\documentclass{aa}

\usepackage{graphicx}
\usepackage[varg]{txfonts}
\usepackage[colorlinks=true,linkcolor=purple,citecolor=purple,urlcolor=blue]{hyperref} 
\usepackage{enumerate}
\usepackage{amsmath}
\usepackage{natbib}
\bibpunct{(}{)}{;}{a}{}{,}
\usepackage{comment}
\usepackage[para]{threeparttable}
\usepackage{multirow}
\usepackage{caption}
\usepackage{floatrow} 
\usepackage{subcaption}
\usepackage{xcolor}
\usepackage{float}
\usepackage{placeins}
\usepackage{afterpage}
\usepackage{array,booktabs}
\begin{document}

\title{High-resolution, 3D radiative transfer modelling \\ III. The DustPedia barred galaxies.}

\author{
Angelos Nersesian\inst{1,2,3} \and
Sam Verstocken\inst{3} \and
S\'{e}bastien Viaene\inst{3,4} \and
Maarten Baes\inst{3} \and
Emmanuel M. Xilouris\inst{1} \and
Simone Bianchi\inst{5} \and
Viviana Casasola\inst{5,6} \and
Christopher J. R. Clark\inst{7} \and 
Jonathan I. Davies\inst{8} \and 
Ilse De Looze\inst{3,9}\and
Pieter De Vis\inst{8} \and
Wouter Dobbels\inst{3} \and
Jacopo Fritz\inst{10} \and
Maud Galametz\inst{11} \and 
Fr\'{e}d\'{e}ric Galliano\inst{11} \and 
Anthony P. Jones\inst{12} \and
Suzanne C. Madden\inst{11} \and
Aleksandr V. Mosenkov\inst{13,14} \and
Ana Tr\v{c}ka\inst{3} \and
Nathalie Ysard\inst{12}
}

\institute{National Observatory of Athens, Institute for Astronomy, Astrophysics, Space Applications and Remote Sensing, Ioannou Metaxa and Vasileos Pavlou GR-15236, Athens, Greece \\ \email{\textcolor{blue}{a.nersesian@noa.gr}} \and  
Department of Astrophysics, Astronomy \& Mechanics, Faculty of Physics, University of Athens, Panepistimiopolis, GR15784 Zografos, Athens, Greece \and 
Sterrenkundig Observatorium Universiteit Gent, Krijgslaan 281 S9, B-9000 Gent, Belgium \and
Centre for Astrophysics Research, University of Hertfordshire, College Lane, Hatfield, AL10 9AB, UK \and
INAF - Osservatorio Astrofisico di Arcetri, Largo E. Fermi 5,I-50125, Florence, Italy \and
INAF - Istituto di Radioastronomia, Via P. Gobetti 101, 4019, Bologna, Italy \and
Space Telescope Science Institute, 3700 San Martin Drive, Baltimore, Maryland, 21218, USA \and
School of Physics and Astronomy, Cardiff University, The Parade, Cardiff CF24 3AA, UK \and 
Department of Physics and Astronomy, University College London, Gower Street, London WC1E 6BT, UK \and
Instituto de Radioastronom\'{i}a y Astrof\'{i}sica, UNAM, Campus Morelia, AP 3-72, 58089 Michoac\'{a}n, Mexico \and
Laboratoire AIM, CEA/DSM - CNRS - Universit\'{e} Paris Diderot, IRFU/Service d’Astrophysique, CEA Saclay, 91191, Gif-sur- Yvette, France \and
Institut d’Astrophysique Spatiale, UMR 8617, CNRS, Universit\'{e} Paris Sud, Universit\'{e} Paris-Saclay, Universit\'{e} Paris Sud, Orsay, F-91405, France \and
Central Astronomical Observatory of RAS, Pulkovskoye Chaussee 65/1, 196140, St. Petersburg, Russia \and 
St. Petersburg State University, Universitetskij Pr. 28, 198504, St. Petersburg, Stary Peterhof, Russia}

\date{Received 25 June 2019 / Accepted  24 July 2019}

\abstract{\textit{Context}. Dust in late-type galaxies in the local Universe is responsible for absorbing approximately one third of the energy emitted by stars. It is often assumed that dust heating is mainly attributable to the absorption of ultraviolet and optical photons emitted by the youngest ($\le 100~$Myr) stars. Consequently, thermal re-emission by dust at far-infrared wavelengths is often linked to the star-formation activity of a galaxy. However, several studies argue that the contribution to dust heating by much older stellar populations might be more significant than previously thought. Advances in radiation transfer simulations finally allow us to actually quantify the heating mechanisms of diffuse dust by the stellar radiation field.

\textit{Aims}. As one of the main goals in the DustPedia project, we have developed a framework to construct detailed 3D stellar and dust radiative transfer models for nearby galaxies. In this study, we analyse the contribution of the different stellar populations to the dust heating in four nearby face-on barred galaxies: NGC~1365, M~83, M~95, and M~100. We aim to quantify the fraction directly related to young stellar populations, both globally and on local scales, and to assess the influence of the bar on the heating fraction.

\textit{Methods}. From 2D images we derive the 3D distributions of stars and dust. To model the complex geometries, we used \textsc{SKIRT}, a state-of-the-art 3D Monte Carlo radiative transfer code designed to self-consistently simulate the absorption, scattering, and thermal re-emission by the dust for arbitrary 3D distributions.

\textit{Results}. We derive global attenuation laws for each galaxy and confirm that galaxies of high specific star-formation rate have shallower attenuation curves and weaker UV bumps. On average, 36.5\% of the bolometric luminosity is absorbed by dust in our galaxy sample. We report a clear effect of the bar structure on the radial profiles of the dust-heating fraction by the young stellar populations, and the dust temperature. We find that the young stellar populations are the main contributors to the dust heating, donating, on average $\sim59\%$ of their luminosity to this purpose throughout the galaxy. This dust-heating fraction drops to $\sim53\%$ in the bar region and $\sim38\%$ in the bulge region where the old stars are the dominant contributors to the dust heating. We also find a strong link between the heating fraction by the young stellar populations and the specific star-formation rate.}

\maketitle
%

\section{Introduction}

Cosmic dust is one of the fundamental ingredients of the interstellar medium (ISM), and considerably affects many physical and chemical processes. Although dust makes up a small fraction of the total mass of a galaxy, it is responsible for the attenuation and reddening effects at ultraviolet (UV) and optical wavelengths \citep{2018ARA&A..56..673G}. In `typical' modern late-type galaxies, dust absorbs roughly 30\% of the total starlight \citep{2002RvMA...15..239P, 2011ApJ...738...89S, 2016A&A...586A..13V, 2018A&A...620A.112B}, and converts this energy to radiation at the mid-infrared (MIR), far-infrared (FIR), and sub-millimetre (submm) wavelengths \citep{1991AJ....101..354S}. Dust emission in those regimes is often used to trace star-formation activity either from MIR and FIR measurements alone \citep{2007ApJ...666..870C, 2015ApJS..219....8C, 2016MNRAS.461..458D}, or from FIR measurements combined with UV and optical data \citep{1998ARA&A..36..189K, 2009ApJ...703.1672K, 2012ARA&A..50..531K}. However, the contribution of the old, more evolved, stars to the dust heating can be non-negligible \citep[e.g.][]{2004ApJS..154..259H,2010ApJ...714.1256C, 2010A&A...518L..65B, 2015MNRAS.448..135B, 2011AJ....142..111B, 2012MNRAS.419.1833B, 2012ApJ...756...40S, 2014A&A...571A..69D, 2017A&A...599A..64V, 2019ApJ...877..140L, 2019A&A...624A..80N} and therefore needs to be taken into account while estimating the star-formation rate (SFR) of a galaxy.  

In the last decade, spectral energy distribution (SED) fitting codes that use Bayesian analysis and cover broader wavelength ranges rose in number and popularity. SED fitting codes can model panchromatic datasets by enforcing an energy conservation between the absorbed starlight and the re-emitted photons by dust, at longer wavelengths. Codes such as \textsc{CIGALE} \citep[Code Investigating GALaxy Emission;][]{2009A&A...507.1793N, 2019A&A...622A.103B} and \textsc{MAGPHYS} \citep{2008MNRAS.388.1595D}, among many others, make use of libraries and templates that account for the stellar and dust emission of galaxies, providing information about the stellar and dust content, and the efficiency of the interstellar radiation field (ISRF) to heat the dust. Despite their rise in popularity, an important caveat of SED fitting still remains: the use of empirical attenuation laws that lack any constraints on the 3D geometry of stars and dust in galaxies. Some codes use different attenuation for young and old stellar populations (SP), but this leads to extra parameters and consequently to degeneracies.

For an accurate and self-consistent representation of the dust-heating processes in galaxies, high-resolution, 3D radiative transfer (RT) modelling is required. Such simulations take into account the complex geometrical distribution of stars and dust (constrained by well-resolved imaging observations), while creating a realistic description of the ISRF as it propagates through the dusty medium. Previous 3D radiative transfer studies have also stressed the importance of non-local dust heating \cite[e.g.][]{2012MNRAS.419..895D, 2014A&A...571A..69D, 2017A&A...599A..64V, 2019MNRAS.487.2753W}, an effect that is not considered in global SED fitting methods or in pixel-by-pixel SED fits. Quantifying the relative contribution of the dust-heating sources through radiation transfer will enable us to obtain an indicative estimation of the current star-forming activity and a better insight on the dust properties in nearby galaxies.

The first detailed radiation transfer models were performed for a slew of edge-on galaxies, using axially symmetric models. Edge-on galaxies offer valuable information on the vertical and radial distribution of their stellar and dust components \citep{1999A&A...344..868X, 2007A&A...471..765B, 2010A&A...518L..39B, 2012MNRAS.427.2797D, 2012MNRAS.419..895D, 2014MNRAS.441..869D, 2015MNRAS.451.1728D, 2016A&A...592A..71M, 2018A&A...616A.120M}. While edge-on galaxies provide significant insight on the vertical and radial structures of galaxies, their main drawback is the lack of insight in the spatial distribution of star-forming regions and the clumpiness of the ISM \citep{2015A&A...576A..31S}. In that sense, well-resolved, face-on galaxies are excellent objects to study since we can identify with great detail the star-forming regions as well as recover the asymmetric stellar and dust geometries. 

A novel technique was developed by \citet{2014A&A...571A..69D} for the panchromatic radiative transfer modelling of well-resolved face-on galaxies. They used observational images to derive the stellar and dust distributions, and then accurately described the starlight-dust interactions. The authors applied this technique to the grand-design spiral galaxy M~51 (NGC~5194). They found that the contribution of the older stellar population to the dust heating is significant, with an average contribution to the total infrared (TIR) emission reaching up to 37\%. \citet{2017A&A...599A..64V} and \citet{2019MNRAS.487.2753W} adopted the same technique and built a radiative transfer model of the Andromeda galaxy (M~31) and the Triangulum Galaxy (M~33), respectively. \citet{2019MNRAS.487.2753W} found that dust in M~33 absorbs 28\% of the energy emitted by the old stellar population, while in M~31 the old stellar population is the dominant dust-heating source with the average contribution being around 91\% \citep{2017A&A...599A..64V}. Furthermore, those three studies have shown that the relative contribution of the young stars to the dust heating varies strongly with location and wavelength. 

Within the scope of the DustPedia\footnote{\url{http://dustpedia.astro.noa.gr}} \citep{2017PASP..129d4102D} project, we have developed a framework to construct detailed 3D panchromatic radiative transfer models for nearby galaxies with the \textsc{SKIRT} Monte Carlo code \citep{2011ApJS..196...22B, 2015A&C.....9...20C}. Where previous works \citep[e.g.][]{2014A&A...571A..69D, 2017A&A...599A..64V, 2019MNRAS.487.2753W} dealt with single individual galaxies, each with their own modelling strategy, here we take advantage of the standardised multi-wavelength imagery data available in the DustPedia archive\footnotemark[1] \citep{2018A&A...609A..37C}, and we apply a uniform strategy for the 3D radiative transfer modelling to a small sample of face-on galaxies with different characteristics. The full description and strategy of our modelling framework is presented in \citet{2019_Verstocken}. The authors applied this state-of-the-art modelling approach to the early-type spiral galaxy M~81 (NGC~3031). In this work we continue this kind of analysis by modelling four late-type barred galaxies; NGC~1365, NGC~5236 (M~83), NGC~3351 (M~95), and NGC~4321 (M~100). Barred galaxies are of particular interest since bars have a strong impact on the physical and chemical evolution of the ISM. It is generally considered that bars funnel molecular gas from the disc toward the central regions of galaxies, fuelling active nuclei and central starbursts \citep[e.g.][]{2011A&A...527A..92C, 2013A&A...558A.124C, 2014A&A...565A..97C}. The 3D distribution of stars and dust in the bars of galaxies could shed light on the physics of the dominant stellar component in both discs and bars. 

\begin{table*}[t]
\caption{Basic properties of the galaxies in our sample.}
\begin{center}
\scalebox{0.87}{
\begin{threeparttable}
\begin{tabular}{lccccccc}
\hline 
\hline 
Galaxy ID & Hubble stage$^\mathrm{(a)}$ & Type$^\mathrm{(b)}$ & Nuclear activity & Distance$^\mathrm{(c)}$ & Apparent size & Position angle & Inclination$^\mathrm{(d)}$\\
& [$T$] & & & [Mpc] & [arcmin] & & \\
\hline
NGC~1365        & 3.2 & SB(s)b  & Seyfert 1.8 & $17.9\pm 2.7$ ($1\arcsec = 86~$pc) & $11.2\times  6.2$ & $132\degr.0$ & $54\degr.5$\\
M~83 (NGC~5236)  & 5.0 & SAB(s)c & Starburst  & $ 7.0\pm 4.1$ ($1\arcsec = 34~$pc) & $12.9\times 11.5$ & $137\degr.0$ & $19\degr.5$\\
M~95 (NGC~3351)  & 3.1 & SB(r)   & Starburst  & $10.1\pm 1.0$ ($1\arcsec = 49~$pc) & $ 7.4\times  5.0$ & $101\degr.2$ & $45\degr.5$\\
M~100 (NGC~4321) & 4.1 & SAB(s)  & H\textsc{ii}/LINER  & $15.9\pm 2.5$ ($1\arcsec = 77~$pc) & $ 7.4\times  6.3$ & $ 84\degr.1$ & $34\degr.9$\\
\hline \hline
\end{tabular}
\begin{tablenotes}
$^\mathrm{(a)}$ From \citet{2014A&A...570A..13M}. $^\mathrm{(b)}$ From \citet{1995yCat.7155....0D}. $^\mathrm{(c)}$ From \citet{2010PASP..122.1397S}. $^\mathrm{(d)}$ From \citet{2019A&A...622A.132M}.
\end{tablenotes}
\end{threeparttable}}
\label{tab:properties}
\end{center}
\end{table*}

This paper is structured as follows. In Sect.~\ref{sec:sample} we present the properties of our galaxy sample as well as a brief description of each galaxy analysed in this study. Section~\ref{sec:model_app} briefly outlines our modelling approach while in Sect.~\ref{sec:mod_valid} we validate our results. In Sect.~\ref{sec:discussion} we show how the contribution of the different stellar populations shapes the SEDs of the galaxies and quantify their contribution to the dust heating. Our main conclusions are summarised in Sect.~\ref{sec:conclusions}. 

\section{Galaxy sample and data} \label{sec:sample}

For the purposes of this work, we selected galaxies from the DustPedia sample \citep{2017PASP..129d4102D} with large angular diameters, so that they are well-resolved even at infrared and submm wavelengths. The sample consists of four nearby, spiral galaxies with a prominent bar in their centres; NGC~1365, M~83, M~95, and M~100. All four galaxies have a small or moderate inclination and optical discs larger than 7~arcmin in diameter. We also selected these galaxies to be roughly representative of early-, mid-, and late-type barred spirals, with the basic properties of each galaxy given in Table~\ref{tab:properties}. For two of them (NGC~1365 and M~83) a detailed analysis of the radial distribution of stars, gas, dust, and SFR is presented in \citet{2017A&A...605A..18C}.

\begin{figure*}[!p]
\centering
    \hfill
    \begin{subfigure}{1.\textwidth}
    \centering
    \includegraphics[width=16.8cm]{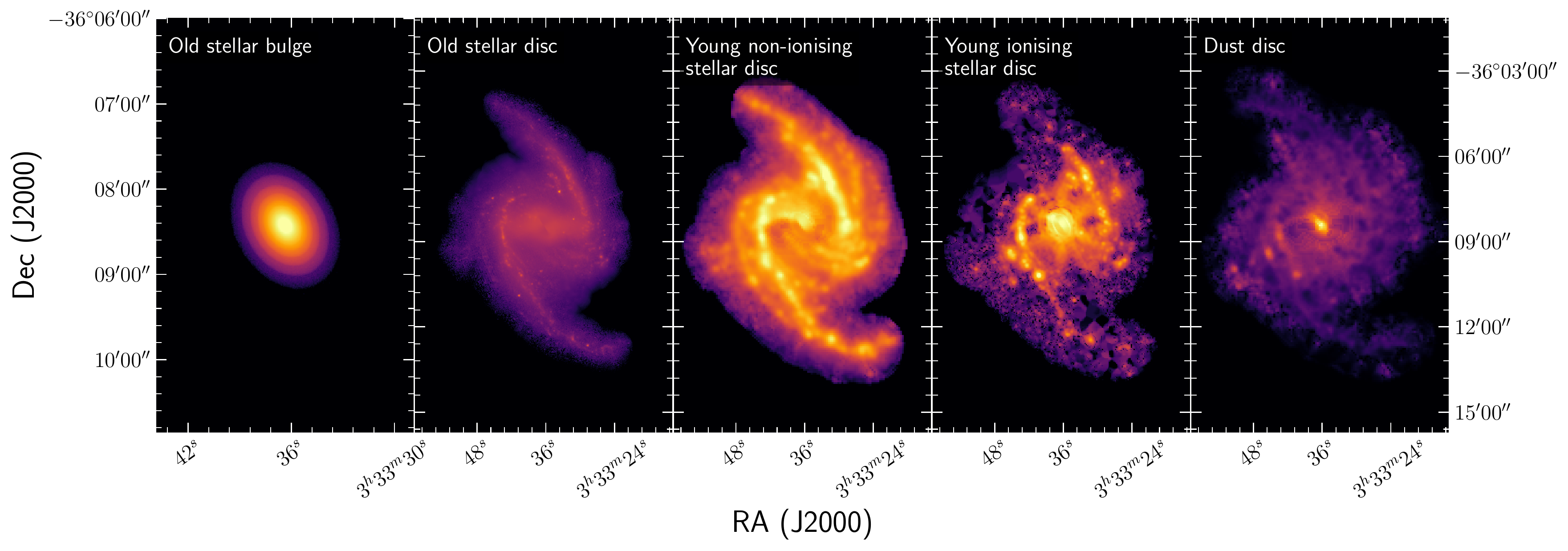}
    \caption{NGC~1365.}\label{subfig:ngc1365_maps}
    \end{subfigure} %
    \hfill
    \begin{subfigure}{1.\textwidth}
    \centering
    \includegraphics[width=16.8cm]{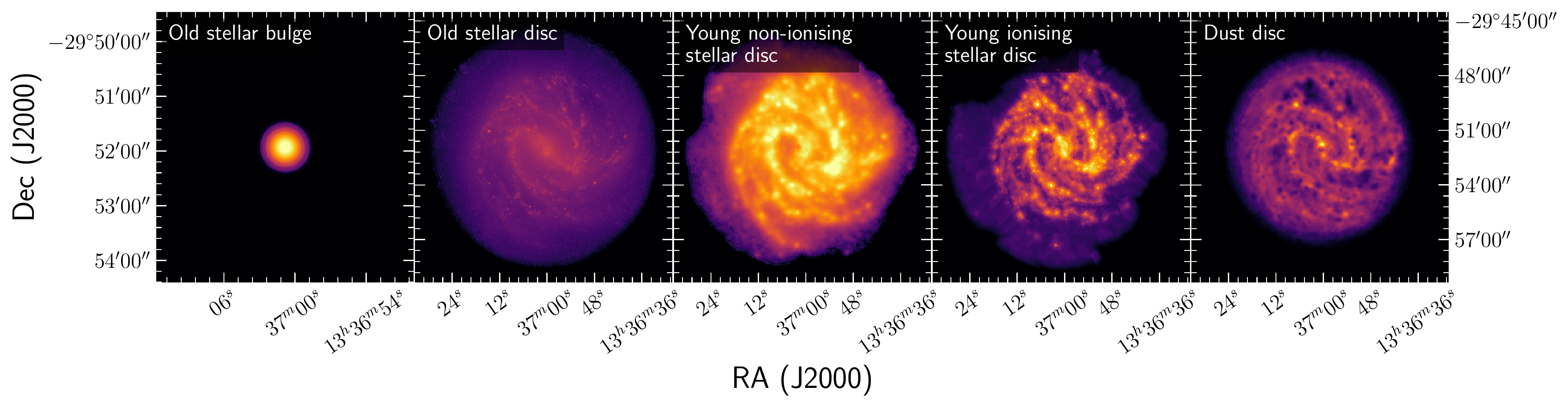}
    \caption{M~83.}\label{subfig:m83_maps}
    \end{subfigure} %
    \hfill
    \begin{subfigure}{1.\textwidth}
    \centering
    \includegraphics[width=16.8cm]{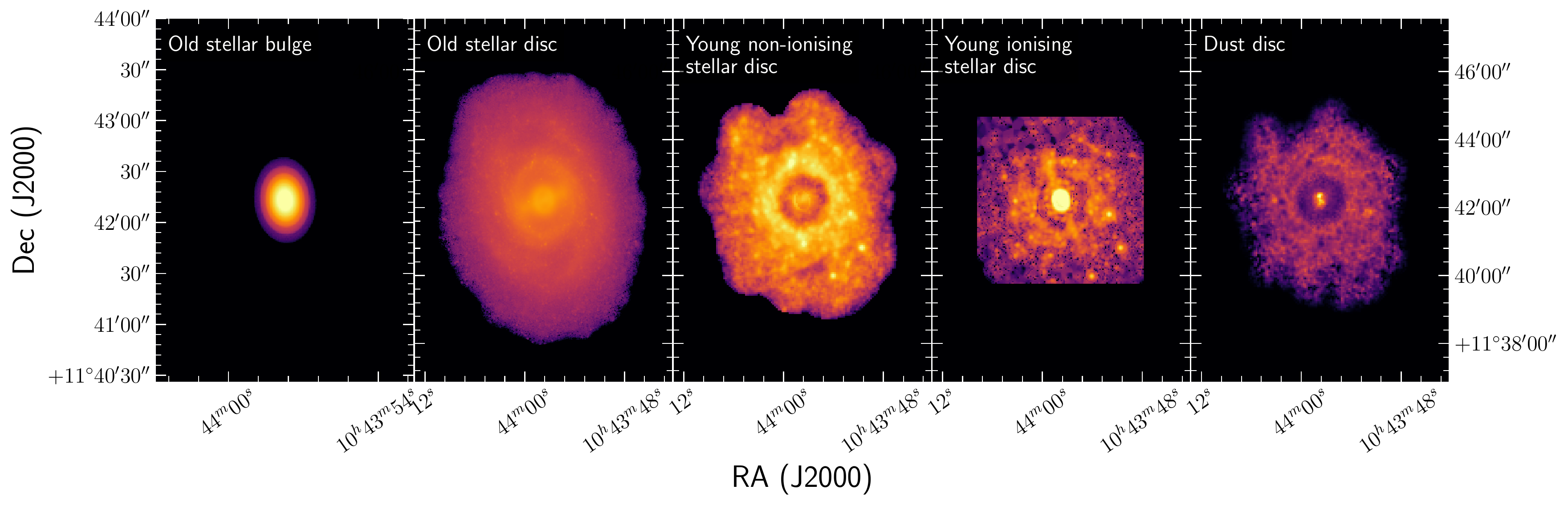}
    \caption{M~95.}\label{subfig:m95_maps}
    \end{subfigure} %
    \hfill
    \begin{subfigure}{1.\textwidth}
    \centering
    \includegraphics[width=16.8cm]{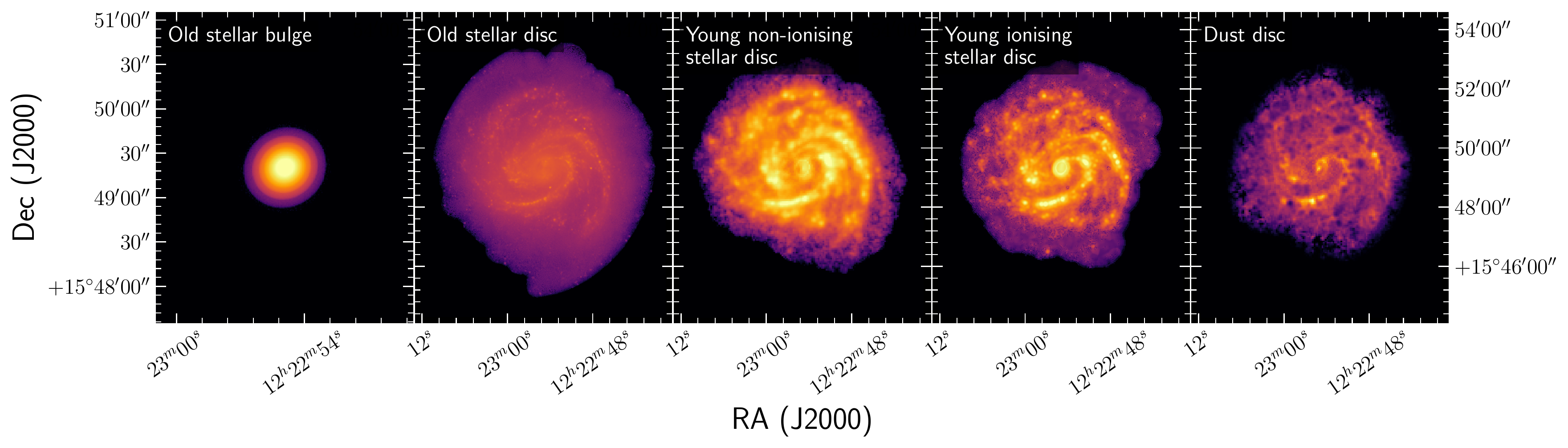}
    \caption{M~100.}\label{subfig:m100_maps}
    \end{subfigure} %
\caption{2D maps of different components of each galaxy. The model includes an old stellar bulge and disc component, a young non-ionising and ionising stellar disc, as well as a dust disc. The bulge image has been generated with \textsc{SKIRT} using a S\'{e}rsic profile geometry. The resolution of each map is based on the respective observations. The extent of the different components is due to the exclusion of unphysical pixels based on the different signal-to-noise ratio (S/N) in each band. The colour coding is in log scale and reflects a normalised flux density.}
\label{fig:maps}
\end{figure*}

NGC~1365 (Fig.~\ref{subfig:ngc1365_maps}), also known as the \textit{Great Barred Spiral Galaxy}, is one of the best studied barred galaxies in the nearby Universe and is located in the Fornax cluster at a distance of 17.9~Mpc \citep{2010PASP..122.1397S}. NGC~1365 has been classified as an SB(s)b type galaxy by \citet{1995yCat.7155....0D} with a Hubble stage of $T=3.2$. This truly impressive galaxy, with a major axis twice as large as of the Milky-Way ($\sim60$~kpc), displays strong ongoing star formation in the centre \citep{1999A&ARv...9..221L} and hosts a bright Seyfert 1.8 nucleus \citep{2006A&A...455..773V}. Two massive prominent dust lanes along the nuclear bar can be seen in optical images \citep{1986MNRAS.221....1T}, while the well developed spiral arms extend from the bar edges with the tendency to turn inwards at the outer edges of the galaxy \citep{1996A&AS..120..403L}. According to \citet{2019A&A...624A..80N}, this massive galaxy contains more than $8 \times 10^{10}~\text{M}_\odot$ of stars, $10^8~\text{M}_\odot$ of dust, and shows a SFR of $13~\text{M}_\odot~\text{yr}^{-1}$. Its H\textsc{i} gas mass is measured to be $9.5 \times 10^9~\text{M}_\odot$ \citep{2019A&A...623A...5D} and its H\textsc{ii} gas mass $3 \times 10^9~\text{M}_\odot$ \citep{2019MNRAS.483.2251Z}.

M~83 (NGC~5236, Fig.~\ref{subfig:m83_maps}) is a grand-design spiral galaxy, with a strong bar in the centre and prominent dust lanes connecting the central region to the disc. M~83 is an SAB(s)c ($T=5$) galaxy as classified by \citep{1995yCat.7155....0D}, representing a `typical' nearby grand-design Sb-Sc galaxy, and is located at a distance of 7.0~Mpc \citep{2010PASP..122.1397S}. It has a nearly face-on orientation, with an estimated inclination of $19\degr.5$. The nuclear region is a site of strong starburst activity \citep{1980ApJ...235..392T, 1987ApJ...313..644T}, with dynamical studies showing that gas is funnelled along the bar producing high rates of star formation at the centre \citep{2010MNRAS.408..797K}. According to \citet{2019A&A...624A..80N}, M~83 has a stellar mass, dust mass, and integrated SFR of $3 \times 10^{10}~\text{M}_\odot$, $2 \times 10^7~\text{M}_\odot$, and $6.7~\text{M}_\odot~\text{yr}^{-1}$, respectively. Its H\textsc{i} gas mass is measured to be more than $2 \times 10^9~\text{M}_\odot$ \citep{2019A&A...623A...5D}. 

M~95 (NGC~3351, Fig.~\ref{subfig:m95_maps}) is a nearby early-type barred spiral galaxy, located at a distance of 10.1~Mpc \citep{2010PASP..122.1397S}. The morphological classification of the galaxy is SB(r) \citep{1995yCat.7155....0D}, with a Hubble stage of $T=3.1$. M~95 is the host of a compact star-forming circumnuclear ring with a diameter approximately 0.7~kpc, and a larger ring of molecular gas regions surrounding the stellar bar of the galaxy \citep{2002MNRAS.337..808K}. Multi-wavelength, sub-kpc studies have shown that the central region is mainly populated by young stars \citep{2013MNRAS.428.2389M}, whereas the bar region mainly hosts an older stellar population \citep{2016MNRAS.457..917J}. According to \citet{2019A&A...624A..80N}, M~95 contains a stellar mass, dust mass, and SFR of $3 \times 10^{10}~\text{M}_\odot$, $8 \times 10^6~\text{M}_\odot$, and $1.1~\text{M}_\odot~\text{yr}^{-1}$, respectively. According to \citet{2019A&A...623A...5D}, this galaxy has an H\textsc{i} gas mass of $10^9~\text{M}_\odot$.

\begin{table*}[ht]
\caption{Overview of the different stellar populations and dust components in our model.}
\begin{center}
\scalebox{0.9}{
\begin{threeparttable}
\begin{tabular}{lllll}
\hline 
\hline 
Component & 2D geometry & Vertical dimension & SED template & Normalisation \\
\hline
\textbf{Bulge} & & & & \\
Old SP (8~Gyr) & \multicolumn{2}{c}{S\'{e}rsic profile geometry$^\mathrm{(a)}$} & \citet{2003MNRAS.344.1000B} & 3.6~$\mu$m \\
\hline
\textbf{Disc} & & Exponential profile & & \\
Old SP (8~Gyr) & IRAC~3.6~$\mu$m$^\mathrm{(b)}$ & $(h_\mathrm{disc,~z})^\mathrm{(f)}$ & \citet{2003MNRAS.344.1000B} & 3.6~$\mu$m\\
Young non-ionising SP (100~Myr) & GALEX~FUV$^\mathrm{(c)}$ & $h_\mathrm{yni,~z}=1/2\times h_\mathrm{disc,~z}$ & \citet{2003MNRAS.344.1000B} & FUV\\
Young ionising SP (10~Myr) & H$\alpha+0.031\times \mathrm{MIPS}~24~\mu$m$^\mathrm{(d)}$ & $h_\mathrm{yi,~z}=1/4 \times h_\mathrm{disc~z}$ & MAPPINGS III$^\mathrm{(g)}$ & FUV\\
Dust & FUV attenuation map$^\mathrm{(e)}$ & $h_\mathrm{dust,~z}=1/2\times h_\mathrm{disc,~z}$ & \textsc{THEMIS}$^\mathrm{(h)}$ dust mix & Total dust mass\\
\hline \hline
\end{tabular}
\begin{tablenotes}
$^\mathrm{(a)}$ The parameters of the flattened S\'{e}rsic profile, like the effective radius $R_\mathrm{e}$, S\'{e}rsic index $n$, and intrinsic flattening factor $q$, were retrieved from the S$^4$G database \citep{2010PASP..122.1397S, 2015ApJS..219....4S}. $^\mathrm{(b)}$ Image corrected for bulge emission. $^\mathrm{(c)}$ Image corrected for old SP emission and dust attenuation \citep{2008MNRAS.386.1157C, 2013MNRAS.431.1956G}, using images from IRAC~3.6~$\mu$m, SDSS~$r$, MIPS~24~$\mu$m, and PACS~70-, 100-, 160$~\mu$m. $^\mathrm{(d)}$ Image corrected for old SP emission. The map was constructed based on the prescription of \citet{2007ApJ...666..870C}. $^\mathrm{(e)}$ The dust map was constructed based on the prescriptions of \citet{2008MNRAS.386.1157C} and \citet{2013MNRAS.431.1956G}. We used images from GALEX~FUV, SDDS~$r$, MIPS~24~$\mu$m, and PACS~70-, 100-, 160$~\mu$m. $^\mathrm{(f)}$ We assumed an exponential distribution with a scale height $h_\mathrm{z}$ in the vertical direction. The scale height for the old SP is, $h_\mathrm{disc,~z} = 1/8.26 \times h_\mathrm{R}$ \citep{2014MNRAS.441..869D}, where $h_\mathrm{R}$ is the scale length. $^\mathrm{(g)}$\citet{2008ApJS..176..438G}. $^\mathrm{(h)}$\citet{2017A&A...602A..46J}.
\end{tablenotes}
\end{threeparttable}}
\label{tab:model_overview}
\end{center}
\end{table*}

M~100 (NGC~4321, Fig.~\ref{subfig:m100_maps}) is located at a distance of 15.9~Mpc \citep{2010PASP..122.1397S} and it is a member of the Virgo Cluster. M~100 has been classified as SAB(s) by \citet{1995yCat.7155....0D}, with two well-defined, symmetrical spiral arms emerging from the bar in the galactic centre. M~100 also hosts a circumnuclear ring with a diameter of 2~kpc. \citet{1997ApJS..112..315H} classified the nucleus as H\textsc{ii}/LINER. The present-day SFR is estimated to be around $6~\text{M}_\odot~\text{yr}^{-1}$ \citep{2019A&A...624A..80N}. M~100 has approximately $5 \times 10^{10}~\text{M}_\odot$ of stars, $4 \times 10^7~\text{M}_\odot$ of dust \citep{2019A&A...624A..80N}, and  $3 \times 10^9~\text{M}_\odot$ of H\textsc{i} gas \citep{2019A&A...623A...5D}.

For all the galaxies in this work, we used imagery data available in the DustPedia archive. These datasets are a combination of images observed by ground-based and space telescopes: GALaxy Evolution eXplorer \citep[GALEX;][]{2007ApJS..173..682M}, Sloan Digital Sky Survey \citep[SDSS;][]{2000AJ....120.1579Y, 2011AJ....142...72E}, 2 Micron All-Sky Survey \citep[2MASS;][]{2006AJ....131.1163S}, Wide-field Infrared Survey Explorer \citep[WISE;][]{2010AJ....140.1868W}, \textit{Spitzer} \citep{2004ApJS..154....1W}; \textit{Herschel} \citep{2010A&A...518L...1P}; and \textit{Planck} \citep{2011A&A...536A..24P}, covering a broad wavelength range from the UV to the submm wavelength domain. For each galaxy, we automatically retrieved more than 24 images from the DustPedia archive through our modelling pipeline \textsc{PTS}\footnote{\url{http://www.skirt.ugent.be/pts8/_p_t_s.html}} \citep[Python Toolkit for \textsc{SKIRT}\footnote{\url{http://www.skirt.ugent.be/root/_landing.html}};][]{2019_Verstocken}. Since SDSS data were not available for NGC~1365 and M~83, we manually downloaded and processed an $R_\mathrm{C}$-band image \citep{1989Msngr..56...31L} for each galaxy from the NASA Extragalactic Database (NED)\footnote{\url{https://ned.ipac.caltech.edu}}. Supplementary to those images, properly processed (stellar continuum-subtracted) H$\alpha$ images were retrieved from NED for all galaxies. 

We pre-processed all images using an automatic procedure as was developed in our modelling framework. First, foreground stars were identified from the 2MASS All-Sky Catalog of Point Sources \citep{2003yCat.2246....0C}, and removed from the GALEX, SDSS, 2MASS, WISE, and \textit{Spitzer} images. Then, all images were corrected for background emission and Galactic extinction. To determine the attenuation in the UV bands we used the \textit{V}-band attenuation, $A_\mathrm{V}$, depending on the position of each galaxy, obtained by querying the IRSA Dust Extinction Service\footnote{\url{https://irsa.ipac.caltech.edu/applications/DUST}}, and by assuming an average extinction law in the Milky-Way (MW), $R_\mathrm{V}=3.1$ \citep{1989ApJ...345..245C}.   

Although multi-wavelength global photometry \citep{2018A&A...609A..37C} is available in the DustPedia archive for all 875 DustPedia galaxies, we performed our own custom aperture photometry using PTS. The reason behind this choice was to ensure that the measurement of the flux densities between observed and simulated images is consistent. The measured flux densities and uncertainties of the observed images are given in Table~\ref{tab:phot}. Throughout the paper we parameterise the galaxy morphology using the Hubble stage ($T$), the values of which have been retrieved from the HyperLEDA database \citep{2014A&A...570A..13M}\footnote{\url{http://leda.univ-lyon1.fr}}. The inclination angle of each galaxy was estimated based on the method described in \citet{2019A&A...622A.132M}. 

\section{Radiative transfer model} \label{sec:model_app}

In this section we briefly lay out the steps we followed to construct our model galaxies. Our purpose is to apply the same systematic approach, introduced in \citet{2019_Verstocken}, for a sample of barred galaxies. For the complete description of our modelling procedure and strategy we refer the reader to \citet{2019_Verstocken}.

\subsection{Radiative transfer simulations with \textsc{SKIRT}}

To generate a 3D radiative transfer model of each galaxy, we used the code \textsc{SKIRT} \citep{2011ApJS..196...22B, 2015A&C.....9...20C}. \textsc{SKIRT} is a radiative transfer code that allows the construction of 3D panchromatic models by using a Monte Carlo approach. The code is designed in a way that it can take into account all relevant physical processes such as scattering, absorption, and thermal re-emission by dust, for a wide variety of environments. \textsc{SKIRT} is equipped with a large collection of possible geometries and geometry decorators \citep{2015A&C....12...33B}, efficient grid structures \citep{2013A&A...554A..10S, 2014A&A...561A..77S, 2013A&A...560A..35C}, and hybrid parallelisation techniques \citep{2017A&C....20...16V}. \citet{2014A&A...571A..69D} implemented a new feature in the code that allows the construction of the complex 3D structures seen in galaxies, from 2D images. The 2D geometry is deprojected and then according to a vertical exponential profile it is smeared out in the vertical direction, so that the flux density is conserved during the conversion from 2D to 3D.

\begin{table*}[t]
\caption{Overview of the model parameters for all four galaxies.}
\begin{center}
\scalebox{1}{
\begin{threeparttable}
\begin{tabular}{c|c|c|c|c|c|c}
\hline 
\hline
 & Description & Parameters & NGC~1365 & M~83 & M~95 & M~100\\
\hline
\multirow{9}{*}{Fixed parameters} & \multirow{3}{*}{S\'{e}rsic parameters$^\mathrm{(a)}$} & $n$ & 0.857 & 0.664 & 0.563 & 0.639 \\
 & & $q$ & 0.577 & 0.897 & 0.256 & 0.795\\
 & & $R_\mathrm{e}$ [pc] & 826 & 236 & 314 & 557 \\
\cline{3-7} 
 & \multirow{4}{*}{Scale heights$^\mathrm{(b)}$} & $h_\mathrm{disc,~z}$ [pc] & 1000 & 436 & 344 & 572\\
 & & $h_\mathrm{yni,~z}$ [pc] & 500 & 218 & 172 & 286\\
 & & $h_\mathrm{yi,~z}$ [pc] & 250 & 109 & 86 & 143\\ 
 & & $h_\mathrm{dust,~z}$ [pc] & 500 & 218 & 172 & 286\\ 
\cline{3-7}  
 & \multirow{2}{*}{Old SP luminosity} & $L_\mathrm{bulge,~3.6}$ [$10^9~\mathrm{L}_\odot$] & 2.60 & 0.68 & 0.35 & 0.67\\
 & & $L_\mathrm{disc,~3.6}$ [$10^9~\mathrm{L}_\odot$] & 7.52 & 7.54 & 1.87 & 5.96\\ 
\hline 
\multirow{6}{*}{Free parameters} & \multirow{3}{*}{Initial guess} & $L_\mathrm{yni,~FUV}^\mathrm{init.}$ [$10^{10}~\mathrm{L}_\odot$] & 3.82 & 4.26 & 0.29 & 1.81\\
 & & $L_\mathrm{yi,~FUV}^\mathrm{init.}$ [$10^{10}~\mathrm{L}_\odot$] & 2.45 & 1.26 & 0.20 & 1.13\\
 & & $M_\mathrm{dust}^\mathrm{init.}$ [$10^7~\mathrm{M}_\odot$] & 10.1 & 2.01 & 0.82 & 3.70\\
\cline{3-7}  
 & \multirow{3}{*}{Best-fit} & $L_\mathrm{yni,~FUV}$ [$10^{10}~\mathrm{L}_\odot$] & $1.21\pm0.70$ & $2.40\pm0.83$ & $0.04\pm0.01$ & $1.02\pm0.37$ \\
 & & $L_\mathrm{yi,~FUV}$ [$10^{10}~\mathrm{L}_\odot$] & $1.83\pm0.60$ & $0.71\pm0.57$ & $0.20\pm0.04$ & $0.36\pm0.23$ \\
 & & $M_\mathrm{dust}$ [$10^7~\mathrm{M}_\odot$] & $18.0\pm4.87$ & $4.76\pm0.98$ & $1.68\pm0.27$ & $6.60\pm1.64$ \\
\hline \hline
\end{tabular}
\begin{tablenotes}
$^\mathrm{(a)}$ Bulge parameters: $n$ is the S\'{e}rsic index, $q$ is the intrinsic flattening factor, and $R_\mathrm{e}$ is the effective radius. $^\mathrm{(b)}$ Disc parameters.
\end{tablenotes}
\end{threeparttable}}
\label{tab:model_params}
\end{center}
\end{table*}

\subsection{Modelling approach}

The model for every galaxy consists of four stellar components and a dust component. We considered an old stellar bulge, an old stellar disc, a young non-ionising stellar disc, a young ionising stellar disc, and a dust disc. We modelled the old and young non-ionising stellar populations using the \citet{2003MNRAS.344.1000B} single stellar population (SSP) templates of solar metallicity $Z = 0.02$, typical ages of 8~Gyr and 100~Myr, respectively, and a \citet{2003PASP..115..763C} initial mass function (IMF). For the young ionising population we adopted the SED templates from MAPPINGS III \citep{2008ApJS..176..438G} assuming an age of 10~Myr. There are five parameters that define the MAPPINGS III templates, namely: the mean cluster mass ($M_\mathrm{cl}$), the gas metallicity ($Z$), the compactness of the clusters ($C$), the pressure of the surrounding ISM ($P_0$), and the covering fraction of the molecular cloud photo-dissociation regions ($f_\mathrm{PDR}$). The following parameters were used as our default values: $Z = 0.02$, $M_\mathrm{cl} = 10^5~\mathrm{M}_\odot$, $\log C = 6$, $P_0/k = 10^6~K~\mathrm{cm}^{-3}$, and $f_\mathrm{PDR} = 0.2$ \citep{2019_Verstocken}. Despite the fact that all four galaxies have a prominent bar in their central region, we did not treat the bar as a separate component here. Instead we treated the bar and the galactic disc as a single structure to keep the modelling procedure in line with the DustPedia standard.

Based on the observed images of the individual galaxies at different wavelengths, we were able to generate the geometrical distribution of each of the input components. We modelled the bulge of each galaxy with a flattened S\'{e}rsic profile. The decomposition parameters of the S\'{e}rsic geometry were retrieved from the S$^4$G database\footnote{\url{https://www.oulu.fi/astronomy/S4G_PIPELINE4/MAIN}} \citep[Spitzer Survey of Stellar Structure in Galaxies:][]{2010PASP..122.1397S, 2015ApJS..219....4S} and we fixed the total luminosity such that it corresponds to the bulge luminosity, measured from the InfraRed Array Camera \citep[IRAC;][]{2004ApJS..154...10F}~3.6~$\mu$m image. To derive the stellar and dust geometries in the disc of each galaxy, we combined different images to create physical maps that characterise, for example, the density distribution of the diffuse dust or old stellar population on the galaxy. The different components can be seen in Fig.~\ref{fig:maps}, while in Table~\ref{tab:model_overview} we provide an overview of the images and templates used for the different stellar and dust components in our model. The details on how we generated these physical maps are presented in \citet{2019_Verstocken}. 

For the dust composition we used the DustPedia reference dust model \textsc{THEMIS}\footnote{\url{https://www.ias.u-psud.fr/themis/THEMIS\_model.html}} \citep[The Heterogeneous Evolution Model for Interstellar Solids;][]{2013A&A...558A..62J, 2017A&A...602A..46J, 2014A&A...565L...9K}. The adopted \textsc{THEMIS} model is for the MW diffuse ISM and even though we know that the dust evolves \citep[e.g.][]{2007ApJ...663..320F, 2009ApJ...699.1209F, 2011A&A...536A...1P, 2011A&A...536A..24P, 2014ApJ...780...10L, 2014ApJ...783...17L, 2015A&A...577A.110Y, 2015ApJ...811..118R, 2017ApJ...834...63R, 2017ApJ...851..119R, 2017ApJ...846...38L, 2018ApJ...862...49N}, this is not taken into account in the current modelling. The dust around star forming regions is introduced in our model through subgrid models that rely on the MAPPINGS III SED templates for the young ionising stellar population (which account for the combined emission from H\textsc{ii} regions and their surrounding PDRs). 

To create the 3D distribution of the disc components we assigned to each of them an exponential profile of different scale heights, $h_\mathrm{z}$, based on previous estimates of the vertical extent of edge-on galaxies \citep{2014MNRAS.441..869D}. Then we generated a dust grid based on the dust component map through which the photons propagate in our simulations. For that purpose, a binary tree dust grid \citep{2014A&A...561A..77S} was employed with approximately 2.8 million dust cells for each galaxy.

Apart from the geometrical distribution, for each stellar component we assigned an intrinsic SED and a total luminosity, that is either fixed or a free parameter in the model. The different combinations of the free parameters generate a 3D radiative transfer simulation that takes into account the emission of the different stellar components as well as the absorption, scattering, and thermal re-emission by dust. The output of each simulation includes the SED of the galaxy and a set of broadband images that can directly be compared to the observed images. Additional information is also available: images of the galaxy at any viewing angles and any wavelengths can be retrieved, and most importantly the effects of the interaction between the ISRF and the diffuse dust can be studied in 3D. Global luminosities are distributed on the 3D pixels (voxels) according to the density distributions as prescribed by the physical maps. The $3.6~\mu$m luminosity of the old stellar population ($L_\mathrm{bulge+disc,~3.6}$) was fixed a priori. 

In Table~\ref{tab:model_params} we list the main parameters that were used to model each galaxy. We distinguish between the parameters that were kept fixed and those that were left free. We left three parameters in our model free and they are determined via a $\chi^2$ optimisation procedure. These parameters are the intrinsic FUV luminosity of the young non-ionising stellar population ($L_\mathrm{yni,~FUV}$), the intrinsic FUV luminosity of the young ionising stellar population ($L_\mathrm{yi,~FUV}$), and the total dust mass ($M_\mathrm{dust}$). In a similar fashion as in \textsc{CIGALE}, we added quadratically an extra 10\% of the observed flux to the measured uncertainty, to account for systematic errors in the photometry and the models \citep{2009A&A...507.1793N}. For the free parameters we provide the initial guess values retrieved from global SED fitting with \textsc{CIGALE}, performed by \citet{2019A&A...624A..80N} for the DustPedia galaxies, as well as the best-fitting values retrieved from our simulations. 

\begin{table}
\caption{Number of grid points for each free parameter ($L_\mathrm{yni,~FUV}$, $L_\mathrm{yi,~FUV}$ and $M_\mathrm{dust}$), and for each batch of simulations. The last column gives the total number of simulations we ran for each galaxy.}
\begin{center}
\scalebox{1.0}{
\begin{tabular}{l|ccc}
\hline 
\hline 
\multirow{2}{*}{Galaxy ID} & \multicolumn{3}{c}{Number of simulations} \\
\cline{2-4}
 & 1st batch & 2nd batch & Total \\
\hline
NGC~1365 & $5\times 7\times 5 = 175$ & $6\times 8\times 6 = 288$ & 463\\
M~83     & $5\times 7\times 5 = 175$ & $7\times 5\times 7 = 245$ & 420\\
M~95     & $5\times 7\times 5 = 175$ & $5\times 5\times 5 = 125$ & 300\\
M~100    & $5\times 7\times 5 = 175$ & $5\times 5\times 5 = 125$ & 300\\
\hline \hline
\end{tabular}}
\label{tab:num_sim}
\end{center}
\end{table}

In order to determine the best-fitting model of each galaxy, we set up two batches of simulations. The first batch acts as an exploratory step of the parameter space. We first generated a broad parameter grid, considering 5 grid points for $L_\mathrm{yni,~FUV}$ and $M_\mathrm{dust}$, and 7 grid points for $L_\mathrm{yi,~FUV}$. The choice of extending the range of $L_\mathrm{yi,~FUV}$ was made because of the difficulty to constrain this particular parameter \citep{2017A&A...599A..64V}. We ran the first batch of simulations with a low-resolution wavelength grid of 115 wavelengths between 0.1 and 1000$~\mu$m, and without the requirement of spectral convolution of the simulated fluxes and images to the filter response curves. For each wavelength we used $10^6$ photon packages, which was sufficient enough to reconstruct the global SEDs. In total, \textsc{SKIRT} created 175 simulated SEDs for each galaxy, and by directly comparing them with the observed SED we were able to narrow down the possible best-fitting parameter ranges. Based on those best-fitting values of the first batch, we generated a refined parameter grid space for the second batch of simulations. 

For the second batch, we used a high-resolution wavelength grid (252 wavelength points) distributed in a non-uniform way over the entire UV-submm wavelength range. Furthermore, we used $5 \times 10^6$ photon packages per wavelength to ensure more accurate sampling of emission, extinction, and scattering, while we enabled spectral convolution. Despite the fact that we started the second batch of simulations with the same number of parameters for all galaxies --5 grid points for each free parameter--, in the cases of NGC~1365 and M~83, the expansion of the parameter space was necessary due to the difficulty in constraining the best-fitting values of $L_\mathrm{yi,~FUV}$ and $L_\mathrm{yni,~FUV}$, respectively. The number of simulations for the first and second batch are given in Table~\ref{tab:num_sim}.

We ran our simulations on the high-performance cluster of Ghent University. For every galaxy here, each simulation of the first batch consumes approximately 22~h of (single-core) CPU time, amounting to 15,400 CPU hours. For the second batch of high-resolution simulations, the average CPU time for each simulation is about 312~h. In total, all simulations together consumed about 26 CPU years. 

\section{Model validation}\label{sec:mod_valid}

\subsection{Global SEDs}\label{subsec:global_sed}

\begin{figure*}[t]
\centering
\includegraphics[width=18cm]{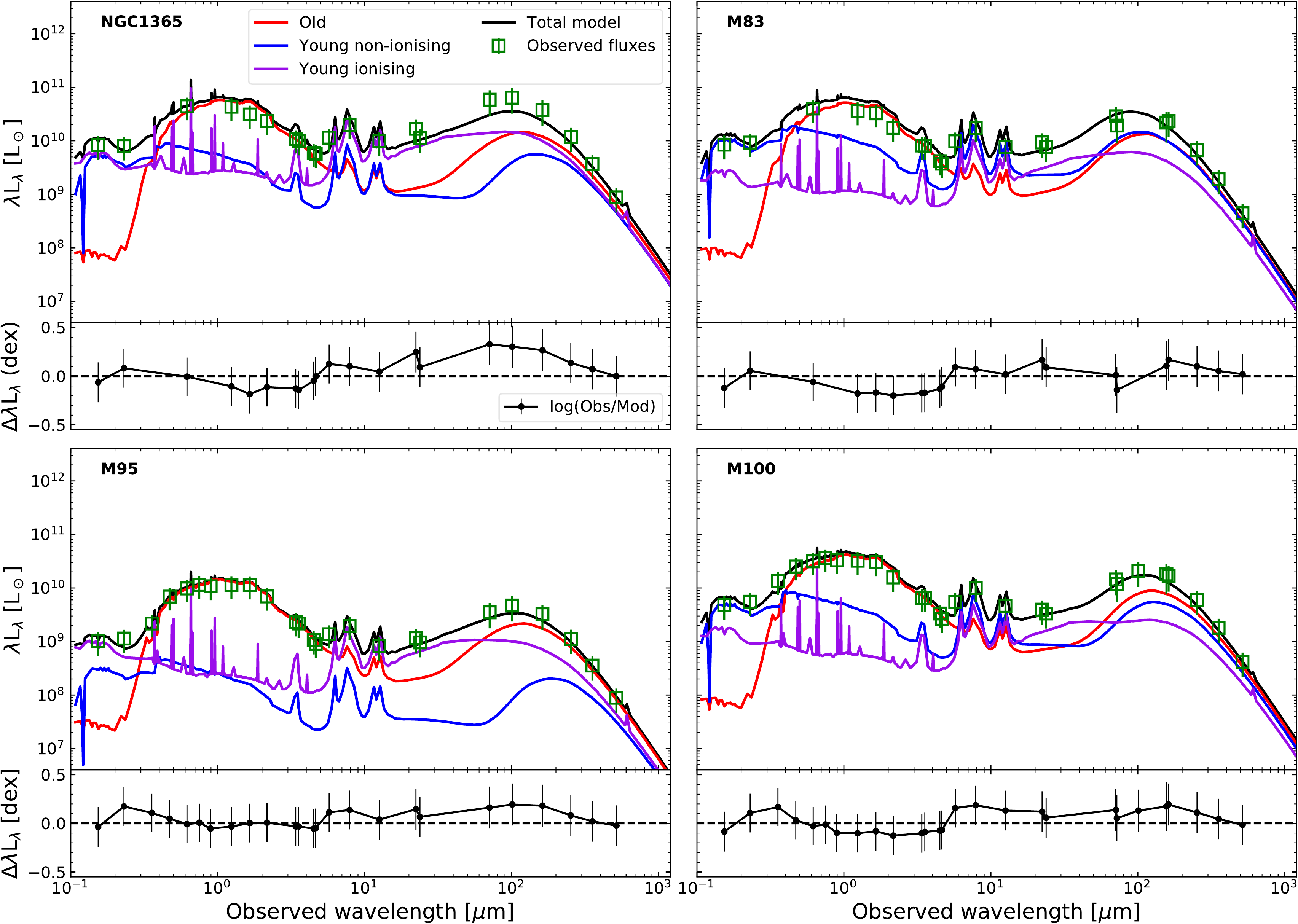}
\caption{The top panel of each sub-figure shows the panchromatic SED of the respective galaxy. The black line is the best-fitting radiative transfer model, run at high-resolution. The green square points are the observed integrated luminosities measured for each galaxy (see Table~\ref{tab:phot}). The red, blue, and violet lines represent the SEDs for simulations with only one stellar component: old, young non-ionising, and young ionising stellar population, respectively. The interstellar dust component is still present in these simulations. The bottom panel of each sub-figure shows the difference in dex between the observations and the best-fitting model.}
\label{fig:seds}
\end{figure*}

We perform a series of quality checks on our results. The best simulated SEDs are shown in Fig.~\ref{fig:seds} along with the observed photometry derived from our pipeline. The total model SED is indicated by the black line, whereas the coloured lines represent the contribution of the different stellar components. We would like to point out that the dust emission of the individual SEDs does not add up to the total SED (black line), because the dependence of dust emission on the absorbed energy is non-linear. On the other hand, the sum of the (attenuated) stellar emission of each component equals the total stellar emission (black line). A weight was assigned to each filter depending on the wavelength regime that it belongs to (we define six regimes: UV, optical, NIR, MIR, FIR, and submm), such that each wavelength regime is equally important. Overall, and in all cases, the simulation agrees with the observations notably well within the uncertainties. 

In all galaxies, a systematic deviation between model and observation is evident for the GALEX~FUV and NUV bands, with the model always overestimating the FUV luminosity and underestimating the NUV luminosity. The most notable differences are around -0.12~dex for the GALEX~FUV band of M~83 and 0.17~dex for the GALEX~NUV band of M~95, while for M~100 there is an equal absolute deviation of 0.1~dex for both wavebands. The discrepancy between model and observation for the UV bands was also reported in the radiation transfer models of M~51 \citep{2014A&A...571A..69D}, M~31 \citep{2017A&A...599A..64V}, M~81 \citep{2019_Verstocken}, and M~33 \citep{2019MNRAS.487.2753W}, as well as for edge-on galaxies \citep[e.g.][]{ 2012MNRAS.427.2797D, 2012MNRAS.419..895D, 2016A&A...592A..71M, 2018A&A...616A.120M}. UV bands are hard to fit because the SED in this spectral region depends sensitively on all the different components: the effects of dust extinction are more pronounced, the shape of the extinction curve is less well-determined and the shape of the intrinsic SEDs is very sensitive to the assumed population ages. Several studies have shown that age-selective attenuation may have a significant effect on the bump strength \citep{1998ApJ...509..103S, 2000ApJ...542..710G} at $0.22~\mu$m, characteristic of the dust attenuation. For example, the MAPPINGS III template seems to induce an inverse UV slope with respect to observations, making it very hard to accurately determine the age of the very young stellar populations. Another and as-yet un-quantified uncertainty arises from \textsc{THEMIS}, which predicts that the UV extinction is sensitive to the a-C nano-particle population \citep{2013A&A...558A..62J} and that this dust component varies with the local ISRF. These effects may result, in part, in the observed discrepancies in the UV bands (especially for the NUV band).

\begin{figure*}[t]
\centering
\includegraphics[width=18cm]{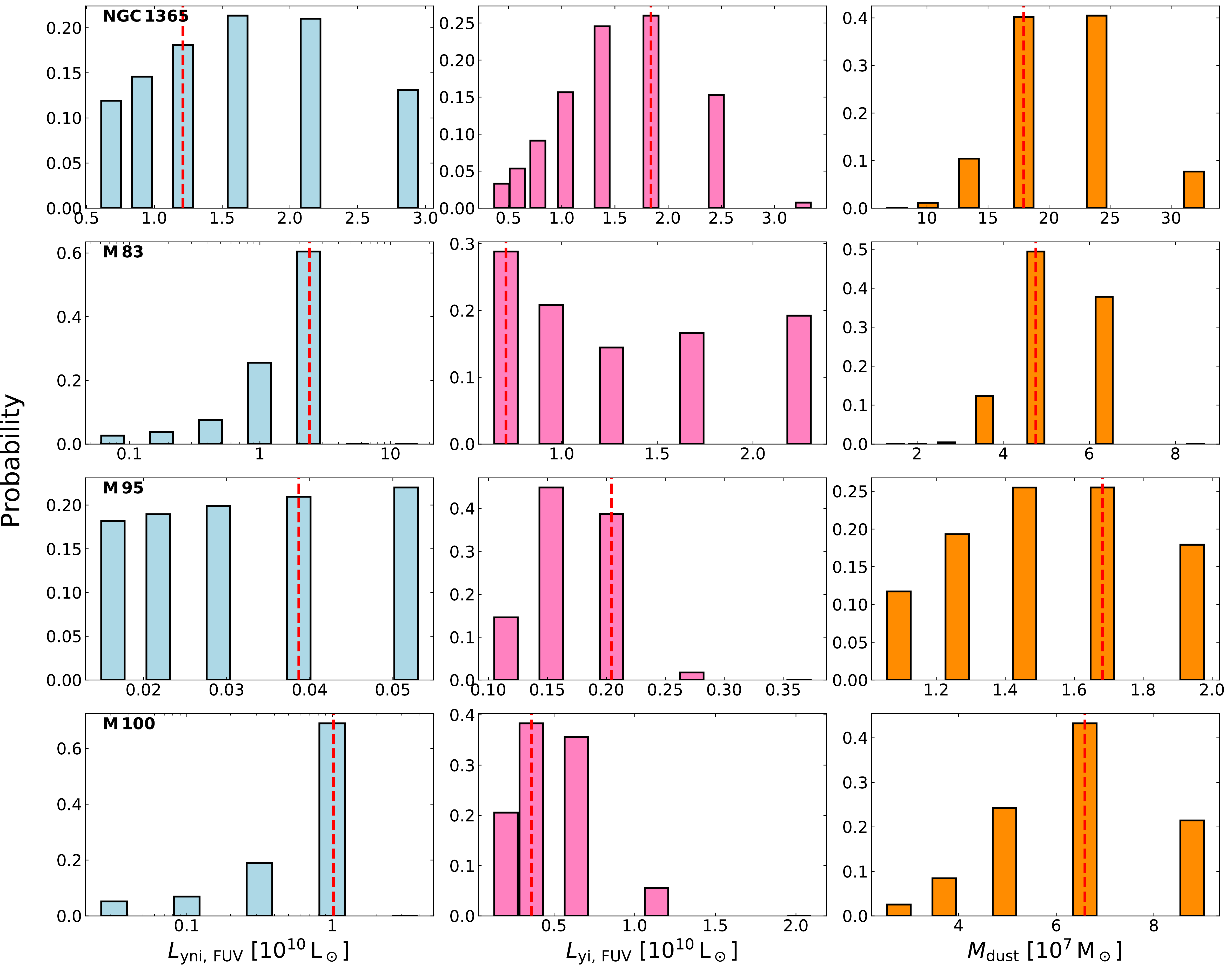}
\caption{PDFs of the three free parameters in our model optimisation: the intrinsic FUV luminosity from the young non-ionising stellar component, $L_\mathrm{yni,~FUV}$ (left), the intrinsic FUV luminosity from the young ionising stellar component, $L_\mathrm{yi,~FUV}$ (middle), and the total dust mass, $M_\mathrm{dust}$ (right). Dashed red lines are the parameter values for the best-fitting model.}
\label{fig:distributions}
\end{figure*}

\begin{figure*}[!p]
\centering
\includegraphics[width=15cm]{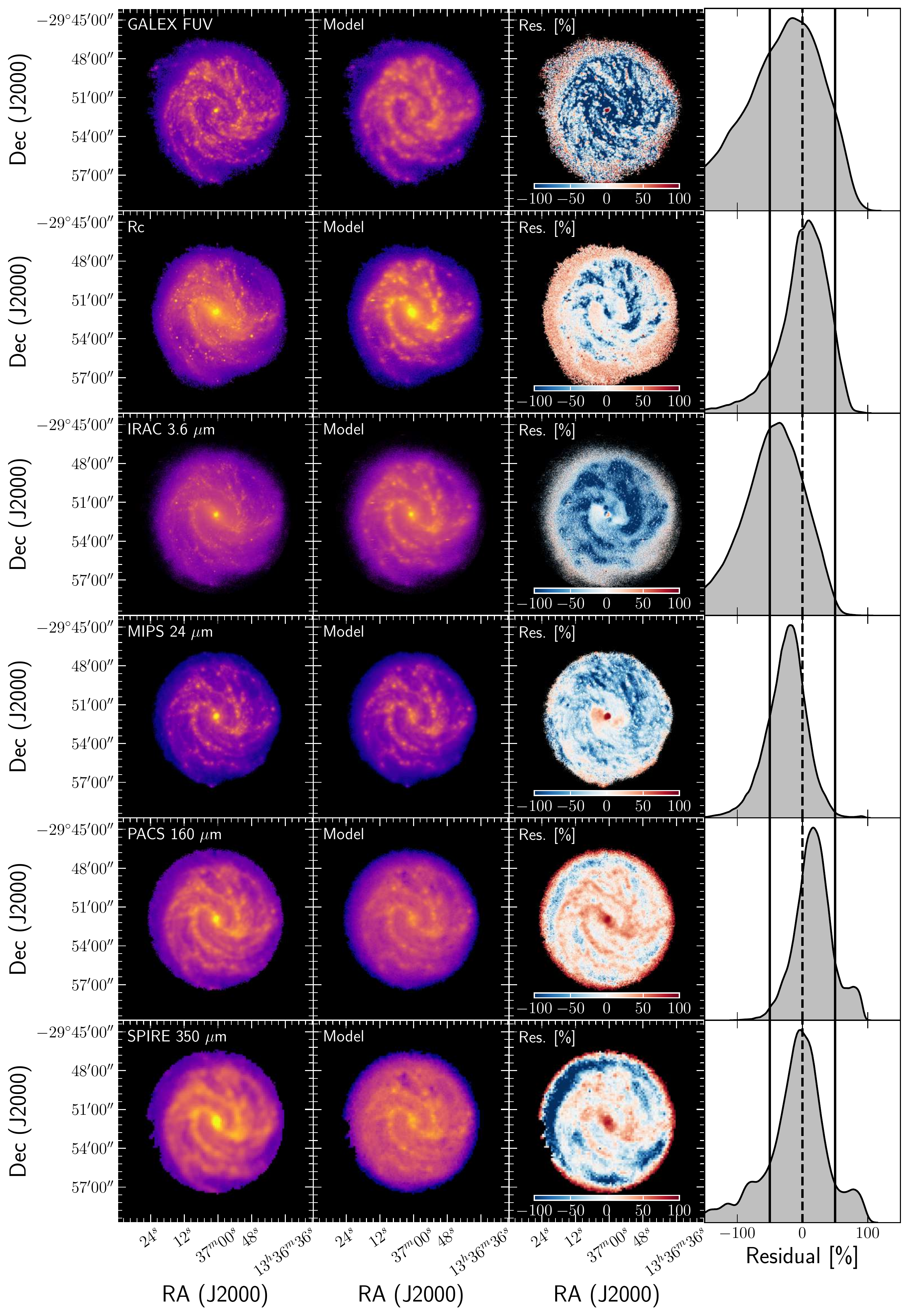}
\caption{Comparison of the simulated images with observations in selected wavebands for M~83. First column shows the observed images, second column the simulated images, third column the maps of the relative residuals between observed and simulated images, and last column shows the KDE of the distributions of the residual pixel values. The simulated images have the same pixel mask as the observed images. The colour coding of the first two columns is in log scale and reflects a normalised flux density. The selected wavebands are: GALEX~FUV, $R_\mathrm{C}$, IRAC~$3.6~\mu$m, MIPS~$24~\mu$m, PACS~$160~\mu$m, and SPIRE~$350~\mu$m.}
\label{fig:res_maps_m83}
\end{figure*}

Another notable discrepancy seen in all galaxies, with the exception of M~95, is the overestimation of the 2MASS bands, with the worst fitted bands being: the 2MASS~$J$ for NGC~1365 with a difference of -0.18~dex, and 2MASS~$K_\mathrm{s}$ for M~83 and M~100 with a difference of -0.2~dex and -0.13~dex, respectively. The 2MASS bands for those three galaxies are less sensitive to more diffuse emission and the 2MASS flux determination is therefore restricted to smaller regions, resulted in lower NIR flux density measurements. Furthermore, even though the predicted MIR-FIR luminosities are fitted by the models within the uncertainties, they fall short in relation to the observations in that wavelength range, especially for NGC~1365, which has a difference for the peak of the dust emission at 100$~\mu$m around 0.3~dex.  

To obtain an estimate of the uncertainty for each one of the free parameters we build their probability distribution functions (PDFs). The probability is proportional to $\exp(-\chi^2/2)$. Figure~\ref{fig:distributions} shows the PDFs of the free parameters from the second batch of simulations for all four galaxies. The best-fitted values are marked by a dashed red line, with the actual values given in Table~\ref{tab:model_params}. In all cases (except the $L_\mathrm{yni,~FUV}$ of NGC~1365), the best-fitted value is either the same as the most probable value of the parameter or it takes the second most probable value. For some parameters an asymmetric distribution is seen, for example in the $L_\mathrm{yni,~FUV}$, of M~83 and M~100 (values of $L_\mathrm{yni,~FUV}$ higher than the best-fitted value were also explored and have close to zero probability). The FUV emission of the young non-ionising stars dominates this particular region of the SED of M~83 and M~100 (blue curve in Fig.~\ref{fig:seds}), leading to a better constraint on the $L_\mathrm{yni,~FUV}$ parameter. On the other hand, the $L_\mathrm{yni,~FUV}$ of M~95 has a flat distribution, leading to a poor constraint on the parameter. The PDFs of the parameters that resemble a normal distribution suggest that the parameters are constrained fairly well. 

Overall, the two fitted luminosities always end up being below the initial guess, while the dust mass always ends up being higher. To be more specific, despite the fact that the best-fitted values of the $L_\mathrm{yi,~FUV}$ are below the initial guess, they are still comparable with the initial guess values (marginally within the uncertainties, with M~100 as the only exception), while the best-fitted values of the $L_\mathrm{yni,~FUV}$ take much lower values than the initial guess. The lower FUV emission from the young non-ionising stellar population could partially explain the lower MIR-FIR emission in the final SED models. The most notable difference between the best-fitted and initial guess values is the total dust mass, where the best-fitted value for all galaxies is approximately two times larger than the one derived by \textsc{CIGALE}. 

\subsection{Image comparison}\label{subsec:OvS}

Another way to validate our results and understand the discrepancies in the integrated luminosities shown in Fig.~\ref{fig:seds}, is to compare model and observations at spatially resolved scales. Figure~\ref{fig:res_maps_m83} shows a selection of representative wavebands across the spectrum of M~83 (similar comparisons for the other three galaxies are provided in the Appendix~\ref{ap:residual_maps}). The bands were selected to demonstrate how well the model reproduces the observed images across the different wavelength regimes (from top to bottom: UV, optical, NIR, MIR, FIR, submm). The most efficient way to visualise any difference between observations (first column) and simulations (second column) is by computing a residual image (third column),

\begin{equation} \label{eq:residuals}
\text{residual} = 100 \times \left(\frac{\text{observation}-\text{model}}{\text{observation}}\right)\% \, .
\end{equation}

\noindent Positive values (in red) mean that the model underestimates the observed emission. On the other hand, negative values (in blue) mean that the model overestimates the observations. The fourth column of Fig.~\ref{fig:res_maps_m83} shows a kernel density estimation (KDE) of the residual values, normalised to 1 at the peak. Overall, there is a good agreement between model and observations for M~83, with the majority of the model pixels within 50\% of their observed counterpart. At this point, we stress that the model images are not directly used in the optimisation procedure (we only fit to the measured global fluxes).

In detail, the FUV emission in the model of M~83 compares quite well with the observations, with the majority of the pixel residuals being near 0\%. However, the model overestimates the FUV emission across the spiral arms and bar region, with residuals lower than -50\%, whereas it underestimates the FUV emission in the central and star-forming regions (red points within the spiral arms). As mentioned in Sect.~\ref{subsec:global_sed}, there is a certain challenge modelling the UV bands because all the different components (stellar and dust) affect, in one way or another, the total emission we observe. 

A very good match between model and observation is seen for the optical image in the $R_\mathrm{C}$ band. The model is an accurate representation of the observed image, with very few residuals below -50\% in the spiral arms, and few positive residuals at the edges of the image due to low S/N. The IRAC~$3.6~\mu$m residual map shows a smooth distribution without many sharp features, with deviations remaining mostly within the spiral arms and partially in the inter-arm regions (i.e. the model overestimates the observations, with the peak of the distribution of the residual values being around 40\%). This somewhat confirms that the old stellar component in our radiation transfer simulations represents the old stellar population adequately. Interestingly enough, the pixel residuals for all three galaxies with high star-formation activity (M~83, NGC~1365, and M~100) display a systematic offset, with the model predicting higher emission despite the fact that we directly determine the normalisation of the old stellar component from the IRAC~$3.6~\mu$m band (see the IRAC~$3.6~\mu$m image comparison in Figs.~\ref{fig:res_maps_m83}, \ref{fig:res_maps_ngc1365}, and \ref{fig:res_maps_m100}). In the 3D model of M~51 \citep{2014A&A...571A..69D}, a similar offset was observed with the model predicting lower values for that band instead. On the other hand, in the 3D models of low star-forming galaxies like M~95 (see the IRAC~$3.6~\mu$m image comparison in Fig.\ref{fig:res_maps_m95}), M~81 \citep{2019_Verstocken}, M~31 \citep{2017A&A...599A..64V}, and M~33 \citep{2019MNRAS.487.2753W}, model and observations are in excellent agreement. The excess $3.6~\mu$m emission may arise from young stars in the spiral arms that contribute light even at these wavelengths. There may also be a contribution from aromatic features emitting at this wavelength. Together these contaminators can explain the differences between model and observations.

Moving on to the MIR, FIR, and submm regimes, simulated images and observations are in good agreement, with residual values having a narrow distribution peaking within $\pm20\%$. The residual map of MIPS~24$~\mu$m exhibits some strong features indicative of the contributions by hot dust and aromatic features of clumpy areas in the outer regions of the spiral arms. In the cases of Photodetector Array Camera and Spectrometer \citep[PACS;][]{2010A&A...518L...2P}~$160~\mu$m and the Spectral and Photometric Imaging REceiver \citep[SPIRE;][]{2010A&A...518L...3G}~$350~\mu$m wavebands, the model underestimates the dust emission mainly in the spiral arms and bar region, with the diffuse dust in the inter-arm regions accurately being represented in our model. 

In addition, when we try to understand the differences between simulations and observations, we need to take into consideration several effects that could potentially increase the level of the observed residuals in Figs.~\ref{fig:res_maps_m83}, \ref{fig:res_maps_ngc1365}, \ref{fig:res_maps_m95}, and \ref{fig:res_maps_m100}. First, the appearance of a significant level of residuals can be attributed to deprojection effects. Due to the deprojection procedure, the brightest regions are smeared out in the direction of deprojection. Then the light is smeared out in the vertical direction creating a blurring-like effect (for example, see the FUV and $R_\mathrm{c}$ model images in Figs.~\ref{fig:res_maps_m83}, \ref{fig:res_maps_ngc1365}, and \ref{fig:res_maps_m100}). Another cause, responsible for a substantial fraction of residuals, is the fact that we combine multiple images, of different resolutions, to generate the input maps of the stellar and dust components in our models. Simulated images thus have a complex point spread function (PSF) and are not convolved by a single beam, in contrast with observed images. 

Finally, a certain degree of difficulty exists when modelling the star-forming regions in detail which probably adds up to the observed discrepancies. For star-forming regions, a spherical shell geometry and an isotropic emission is implemented here. Those models result in a higher level of attenuation per unit dust mass than other models where a clumpy or asymmetric geometry is assumed \citep{1996ApJ...463..681W, 2000ApJ...528..799W, 1999ApJ...523..265V, 2006ApJ...636..362I, 2011ApJ...729..111W}. Of course the effect described here is more pronounced in the UV regime, where the young ionising stars are relatively more luminous. 

\section{Discussion} \label{sec:discussion}

\subsection{Attenuation law}\label{subsec:att_laws}

\begin{figure}[t]
\centering
\includegraphics[width=9cm]{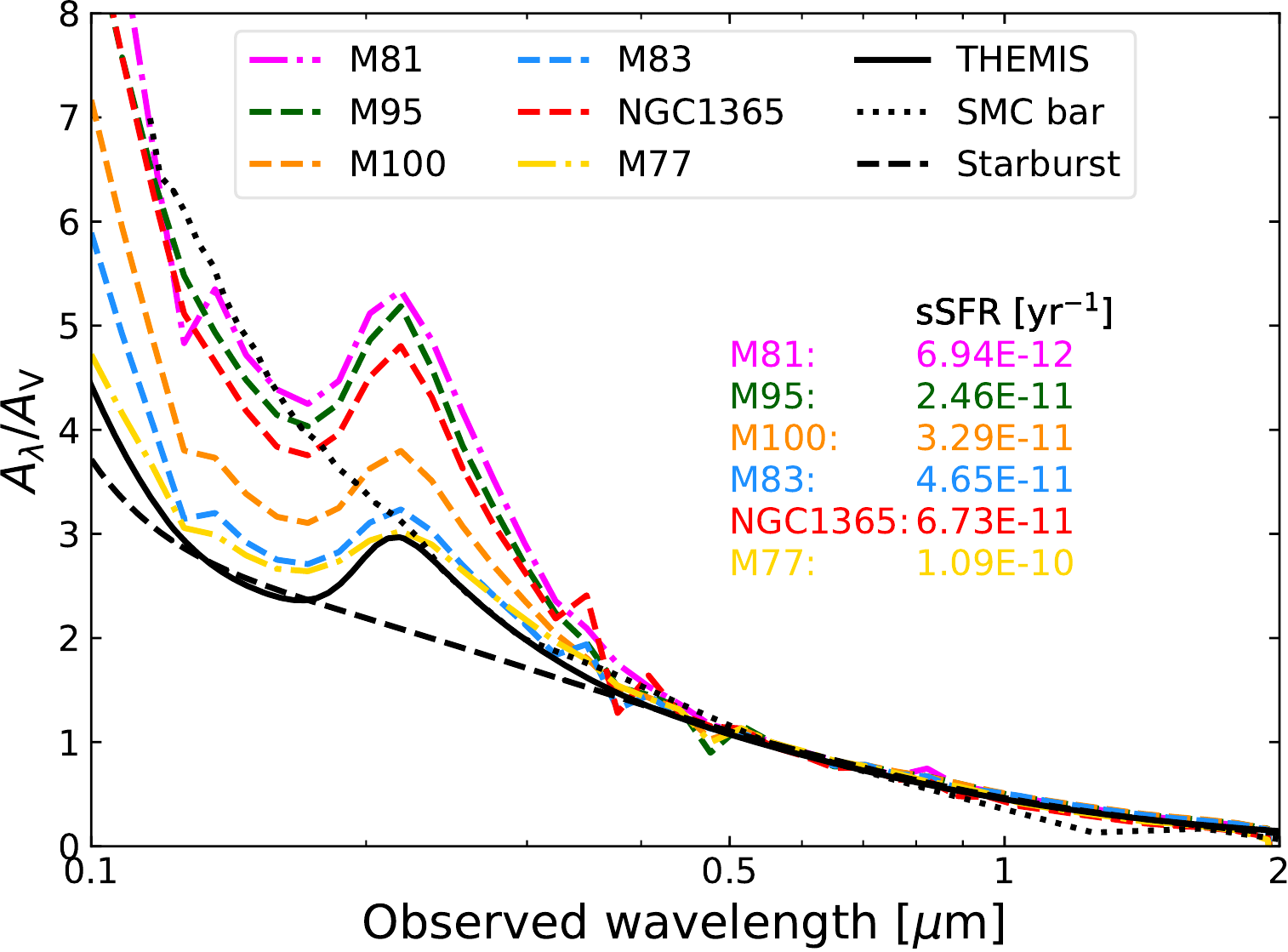}
\caption{Compilation of attenuation laws derived from our models on a face-on view (0$\degr$ inclination angle), normalised to the V-band attenuation. In addition to the attenuation curves derived for the galaxy sample of this study, we provide the attenuation curves of M~81 from \citet{2019_Verstocken}, and M~77 from \citet{Viaene2020}. Several literature measurements are shown as well: solid black curve: \textsc{THEMIS} extinction law \citep{2013A&A...558A..62J, 2017A&A...602A..46J, 2014A&A...565L...9K}, dotted black line: extinction curve of SMC bar \citep{2003ApJ...594..279G}, dashed black line: attenuation law of starburst galaxies \citep{2000ApJ...533..682C}. The median values of the sSFR for every galaxy is also given in increasing order, as derived from our simulations.}
\label{fig:att_curves}
\end{figure}

An important caveat in SED fitting codes is the use of idealised attenuation curves, converted from extinction laws that do not fully incorporate the effect of the relative geometries expected to be found between dust and stars. In a recent effort to address this caveat, \citet{2018A&A...619A.135B} measured the shape of the attenuation curves of star-forming galaxies by employing two different recipes: a flexible Calzetti attenuation law \citep{2009A&A...507.1793N} and a two power-law recipe based on the one inferred by \citet{2000ApJ...539..718C}. Both recipes take the shape of the attenuation curve and the relative attenuation of young and old stellar populations as free parameters. \citet{2018A&A...619A.135B} found that the \citet{2000ApJ...539..718C} recipe is able to better reproduce the results from radiative transfer models, and \citet{2019arXiv190209435B} proposed a new modified Calzetti attenuation law with that specific goal in mind. From our radiative transfer simulations we can shed light on the impact of the relative geometry between the different stellar populations and the diffuse dust to the observed galaxy SEDs by reconstructing realistic dust attenuation curves. In order to determine the global attenuation curves we use the observed SED of the best-fitting model and the stellar spectrum we reconstruct for each galaxy. In Fig.~\ref{fig:att_curves} we present the attenuation curves of the galaxies in our sample, derived from the 3D modelling with \textsc{SKIRT}. A face-on orientation (inclination angle 0$\degr$) was assumed for all galaxies. We complement the attenuation curves with those derived for M~81 from \citet{2019_Verstocken} and M~77 from \citet{Viaene2020}. The attenuation curves of NGC~1365, M~83, M~95, M~100, M~81, and M~77 have been normalised to the $V$-band attenuation by: 0.27, 0.47, 0.22, 0.32, 0.09, and 1.03, respectively. We find a steep increase of attenuation towards the UV wavelengths due to absorption by small grains. A broad absorption bump is also evident with a peak around 0.22$~\mu$m. The values of the $V$-band attenuation indicate that the galactic discs are optically thin if galaxies were to be seen face-on. This explains the steeper slopes in the UV wavelengths and the stronger 0.22~$\mu$m bumps in relation to the normalised \textsc{THEMIS} extinction curve \citep{2000ApJ...528..799W}.

The observed curves are the combination of the attenuation by diffuse dust as modelled by the dust mass input map, and the attenuation by dust in star-forming regions. The dust in the star-forming clouds is incorporated in the MAPPINGS III SED templates \citep{2008ApJS..176..438G}, which were used to represent the ionising stellar component. The MAPPINGS III attenuation law is directly provided by \citet{2008ApJS..176..438G}, while the diffuse dust attenuation curve is a combination of the \textsc{THEMIS} extinction law \citep{2013A&A...558A..62J, 2017A&A...602A..46J, 2014A&A...565L...9K} and the relative geometry between stars and dust. In Fig.~\ref{fig:att_curves}, we also plot the extinction curve of the \textsc{THEMIS} dust model (solid black line), which was used in all six studied galaxies. In addition, for comparison purposes, we provide literature measurements of the extinction law for the Small Magellanic Cloud (SMC) bar region from \citet{2003ApJ...594..279G} (dotted black line), and the attenuation law of starburst galaxies from \citet{2000ApJ...533..682C} (dashed black line). Moreover, we show the median values of the sSFR for every galaxy in increasing order derived from our simulations (see Sect.~\ref{subsec:ssfr} on how we compute the sSFR).

At optical and NIR wavelengths ($\lambda \ge 0.4~\mu$m) all attenuation curves are in agreement, but this is expected since we normalised them with the $V$-band. At shorter wavelengths ($\lambda < 0.4~\mu$m), the curves begin to diverge. The UV bump in all six galaxies covers quite an extended range, with the peak of the bump varying over 2 orders of magnitude, despite the fact that all galaxies in our collective study share the exact same grain properties (i.e. the \textsc{THEMIS} extinction curve for the standard MW case was used in every model). The diversity of the bump strength can be linked directly to the sSFR (a measure of the current to past star formation in galaxies) of each system \citep{2013ApJ...775L..16K, 2015ApJ...806..259R}. A weakened UV bump implies that an extra amount of radiation is filling in the bump either by UV photons from unobscured young stars or by light scattered into the line of sight \citep{2018ApJ...869...70N}. Of course, a reduced UV bump can also arise from opposing processes: intense UV radiation destroying the carriers in low density regions, and the bump carriers accreting onto big grains in high density regions, however these processes are not incorporated in the model so they cannot explain the differences in the attenuation curves. 

With the exception of NGC~1365, we find a correlation between the sSFR and the shape of the attenuation curve, suggesting an age-dependent extinction curve. M~81 and M~95, two low sSFR galaxies exhibit a strong, almost identical UV bump, whereas galaxies of higher sSFR, for example M~83 and M~77, have a weaker UV bump. Contrarily, NGC~1365, a galaxy of high sSFR, presents a strong bump similar to M~81 and M~95. This behaviour suggests that most of the UV light emitted in the star-forming regions of NGC~1365 is heavily obscured by dust resulting in the strong bump feature we observe here. Another notable result is the presence of a UV bump in all attenuation curves, despite the claims made by some other authors in the past that bump-free attenuation curves, such as the \citet{2000ApJ...533..682C} curve, could arise even with dust that has a normal UV bump in the extinction curve. We confirm that as long as the UV bump in the extinction curve is represented as a true absorption feature, the corresponding attenuation laws must have a bump, although it may appear weakened.

Beyond the UV bump, our model curves steadily increase and fall somewhere between the starburst idealised attenuation law and the SMC extinction curve. It is interesting to notice here that stronger bumps seem to sit on steeper slopes. This result is well known and is attributed to the star-dust geometry \citep{1996ApJ...463..681W, 2018ApJ...869...70N}. \citet{2018ApJ...869...70N} have shown that steeper slopes may arise either by a large fraction of obscured young stars or by a significant fraction of unobscured old stars. Consequently, galaxies with older stellar populations exhibit steeper attenuation curves. On the other hand, flatter attenuation curves are the result of a more complex geometry where more of the starlight is decoupled from dust. Based on our results we hereby confirm that galaxies of high sSFR values have shallower attenuation curves and weaker UV bumps.

\subsection{Dust heating}\label{subsec:dust_heating}

A relevant quantity that holds information about the stellar energy absorbed by dust in a galaxy, and that is based on the assumption of energy balance, is the fraction of dust to bolometric luminosity (dust-heating fraction). \citet{2018A&A...620A.112B} defined this quantity as $f_\text{abs}$. With the use of radiative transfer modelling we can calculate this quantity not only on global scales but also on local scales. On global scales the dust-heating fraction is:

\begin{equation} \label{eq:fabs}
f_\text{abs}^\text{SKIRT} = \frac{L_\text{dust}}{L_\text{stars} + L_\text{dust}} \, ,
\end{equation}

\noindent where $L_\text{stars}$ is the observed stellar emission and $L_\text{dust}$ is the total dust luminosity, computed by integrating the SEDs presented in Fig.~\ref{fig:seds}. For each galaxy we have calculated $f_\text{abs}^\text{SKIRT}$ and we have compared them with $f_\text{abs}^\text{CIGALE}$ produced in \citet{2018A&A...620A.112B} (see Table~\ref{tab:fractions}). The $f_\text{abs}^\text{SKIRT}$ fractions we obtain are slightly lower than, but compatible with, those obtained by \citet{2018A&A...620A.112B}. In \citet{2018A&A...620A.112B} the authors provide the mean $f_\text{abs}^\text{CIGALE}$ values for 814 DustPedia galaxies, divided into 6 groups, according to their morphology classification (Hubble stage, $T$). The mean $f_\text{abs}^\text{CIGALE}$ value of the corresponding morphological bin $\left(\text{Sb-Sc}; 2.5 \le T <5.5\right)$ that our galaxy sample falls into is $32.8 \pm 13.9\%$. This value is in very good agreement with the mean value of $f_\text{abs}^\text{SKIRT}$ ($36.5 \pm 7.4\%$) of the rather small group of galaxies in our study. In any case, our modelling gives us the opportunity to better characterise which stellar population is the dominant dust-heating source in each galaxy, and how significant the contribution of old stars is, on a spatially resolved manner.

From our simulations it is possible to retrieve the absorbed energy in each dust cell (originating from the different stellar populations in the model), and thus to quantify the dust-heating fraction from the young non-ionising and young ionising stellar populations (hereafter, we will refer to the heating fraction by the young non-ionising and young ionising stellar populations simply as young heating fraction or $f_\text{young}$). We obtain the young heating fraction through:

\begin{equation} \label{eq:heating_fraction}
f_\text{young} = \frac{L_\text{yni}^\text{abs} + L_\text{yi}^\text{abs}}{L_\text{total}^\text{abs}} \, .
\end{equation}

\noindent where $L_\text{yni}^\text{abs}$ and $L_\text{yi}^\text{abs}$ are the luminosities of the young non-ionising and young ionising stellar populations absorbed by dust, respectively, and $L_\text{total}^\text{abs}$ is the total stellar luminosity absorbed by dust. 

\begin{figure*}[t]
\captionsetup[subfigure]{labelformat=empty}
\centering
    \hfill %
    \begin{subfigure}[b]{0.49\textwidth}
    \centering
    \includegraphics[width=9cm]{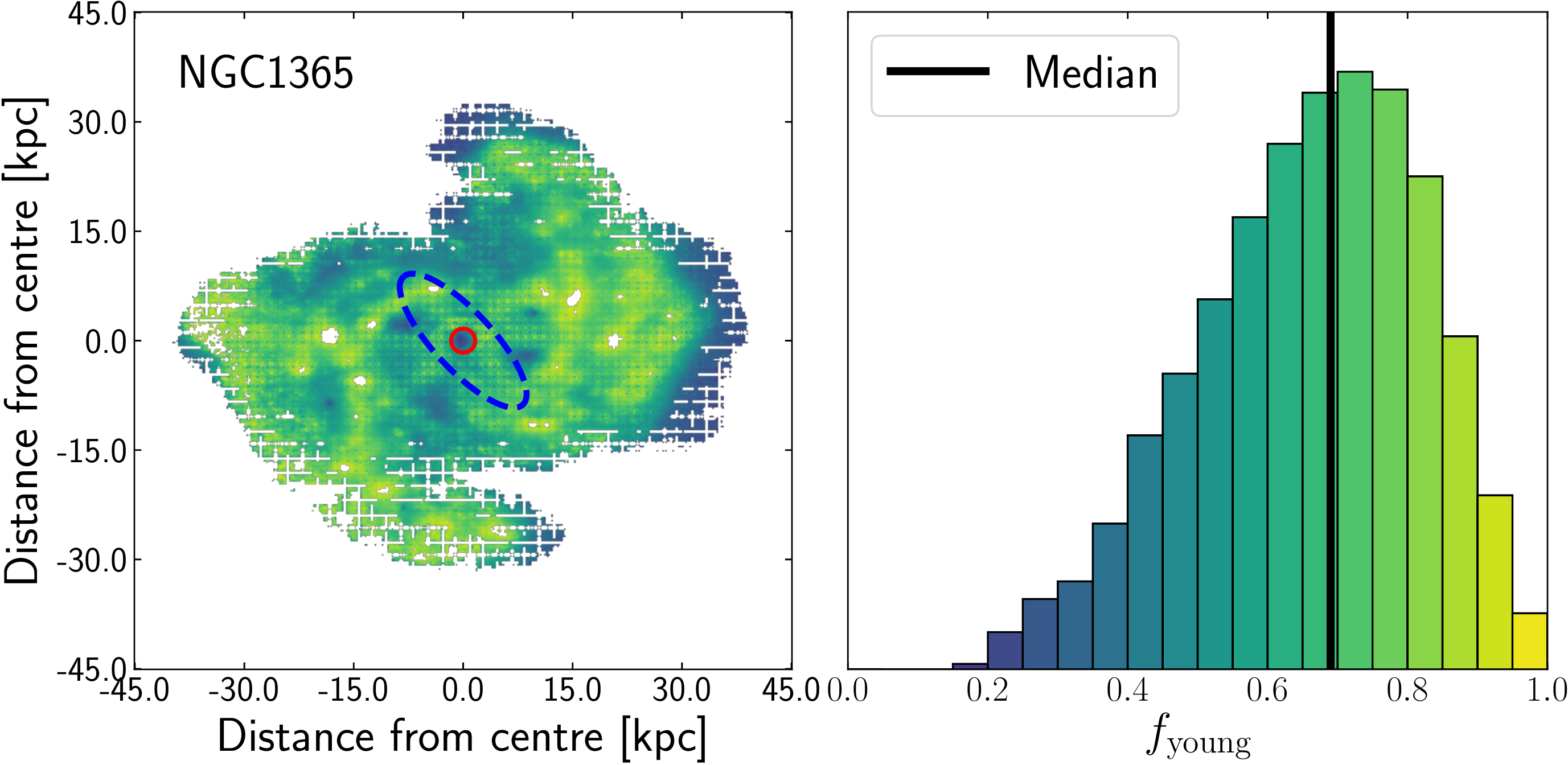}
    \end{subfigure} %
    \hfill %
    \begin{subfigure}[b]{0.49\textwidth}
    \centering
    \includegraphics[width=9cm]{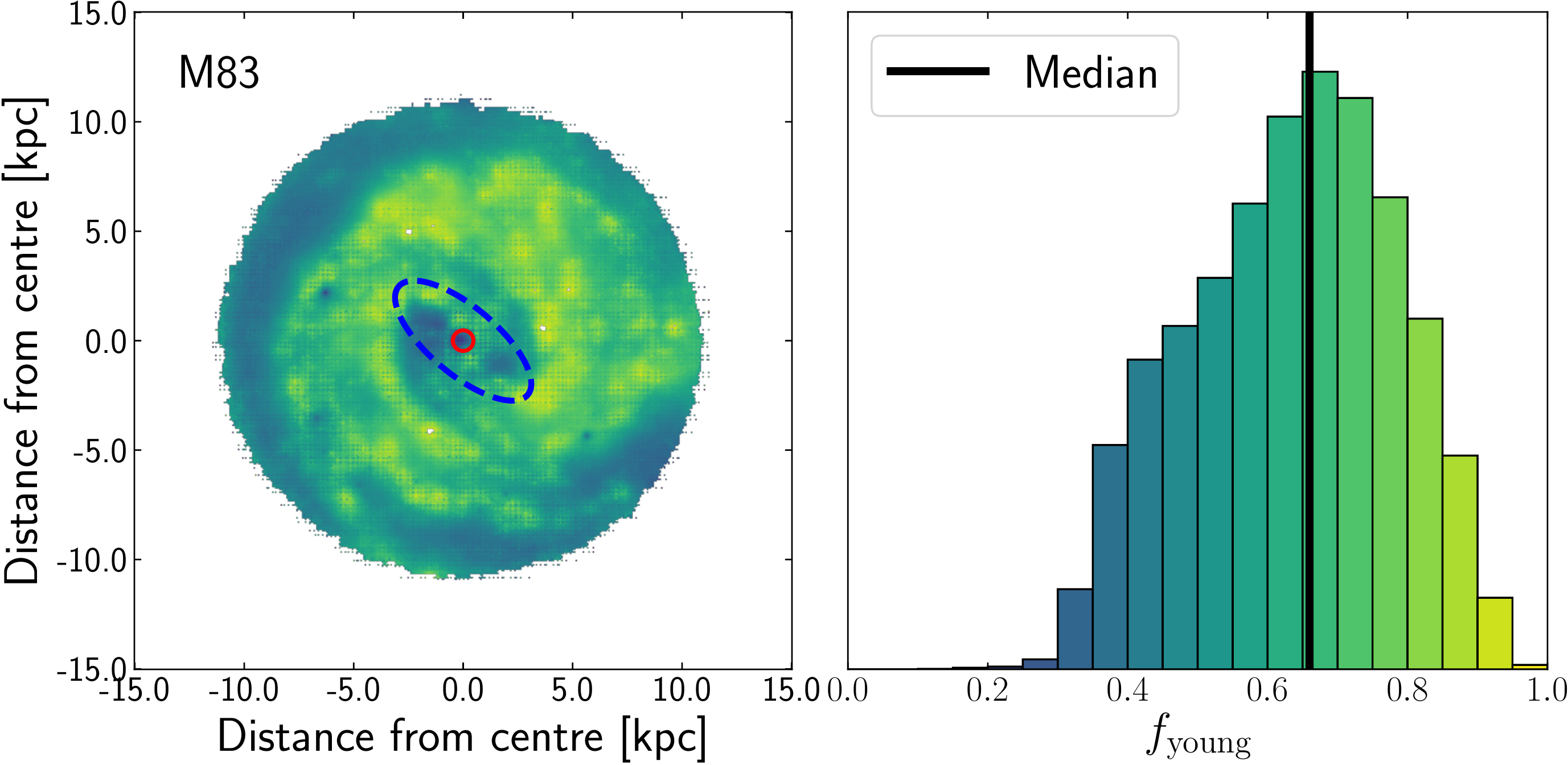}
    \end{subfigure} %
    \hfill %
    \begin{subfigure}[b]{0.49\textwidth}
    \centering
    \includegraphics[width=9cm]{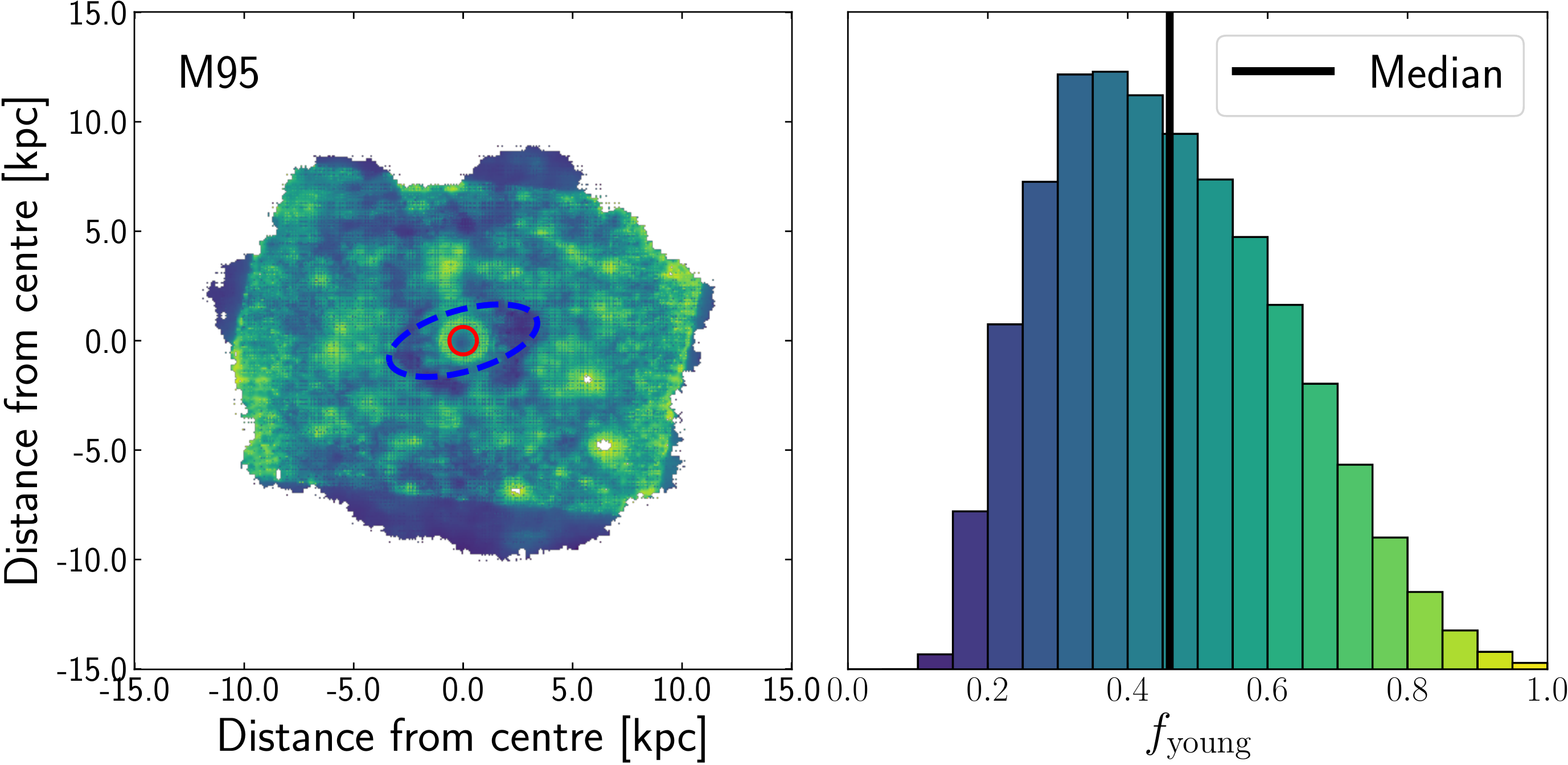}
    \end{subfigure} %
    \hfill %
    \begin{subfigure}[b]{0.49\textwidth}
    \centering
    \includegraphics[width=9cm]{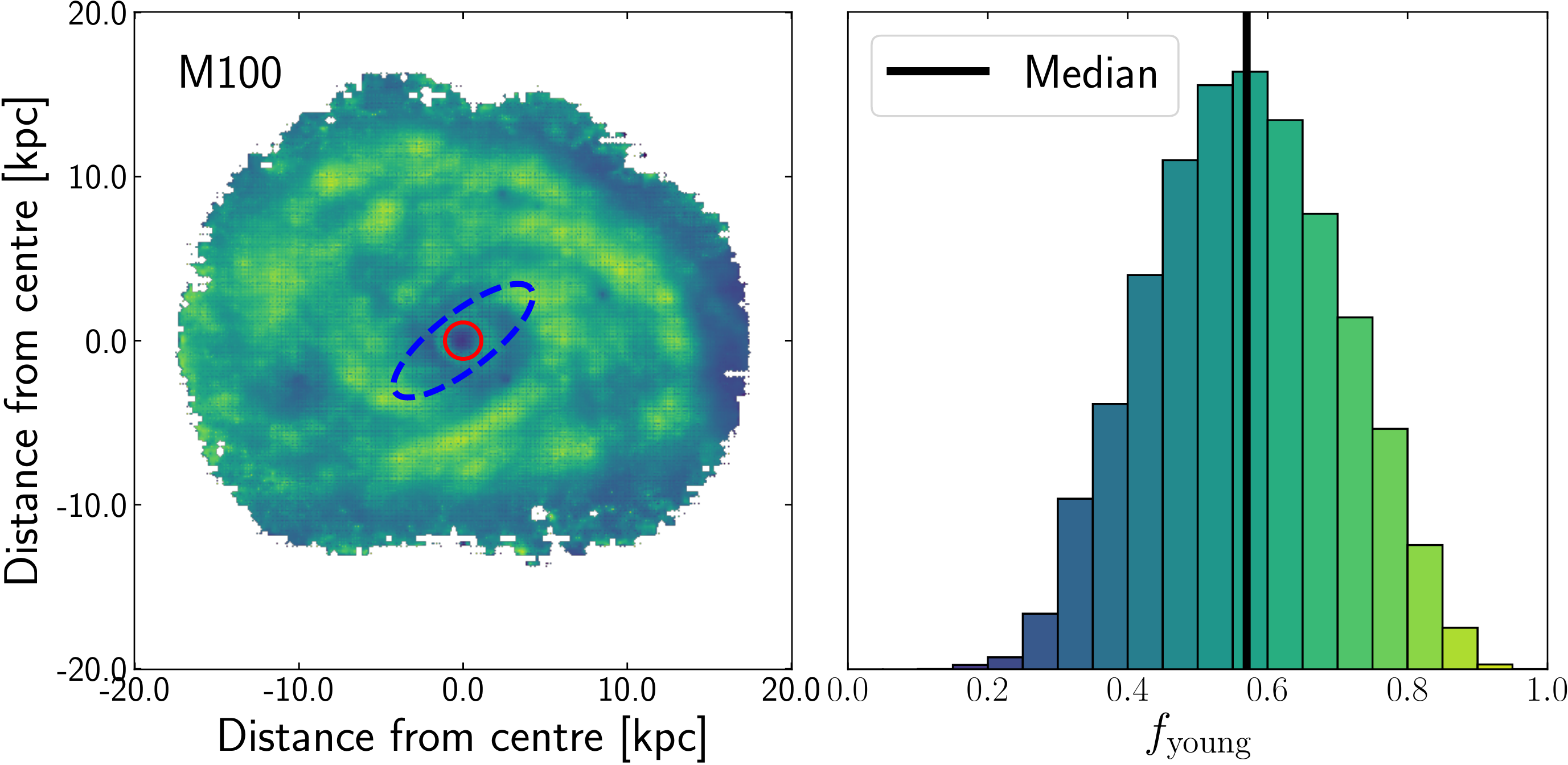}
    \end{subfigure} %
\RawCaption{\caption{Left panel of each sub-figure: Face-on view of the heating fraction by young non-ionising and young ionising stellar populations (Equation~\ref{eq:heating_fraction}), as obtained from the 3D dust cell distribution, for each galaxy. The bulge region of every galaxy is indicated with a solid red circle and the bar region with a dashed blue ellipse (see text for more details). Right panel of each sub-figure: Distribution of the dust cell heating fractions, weighted by dust mass. The histogram also denotes the colour coding of the map on the left. The solid black line shows the median value.}
\label{fig:heating_maps}}
\end{figure*}

\begin{table}
\caption{$f_\text{abs}^\text{SKIRT}$, $f_\text{abs}^\text{CIGALE}$ \citep{2018A&A...620A.112B}, and the mean global $f_\text{young}$, as well as the mean $f_\text{young}$ in the bulge and bar region; the regions are defined in Fig.~\ref{fig:heating_maps} with a solid red circle and a dashed blue ellipse, respectively.}
\begin{center}
\scalebox{0.9}{
\begin{tabular}{l|cc|ccc}
\hline 
\hline 
\multirow{2}{*}{Galaxy ID} & $f_\text{abs}^\text{SKIRT}$ & $f_\text{abs}^\text{CIGALE}$ & \multicolumn{3}{c}{$f_\text{young}$ [\%]} \\
\cline{4-6}
 & [\%] & [\%] & Global & Bulge & Bar \\
\hline
NGC~1365 & 44 & $53\pm21$ & $68\pm15$ & $40\pm14$ & $66\pm13$ \\
M~83     & 43 & $53\pm9$  & $64\pm15$ & $32\pm12$ & $59\pm12$ \\
M~95     & 26 & $28\pm4$  & $47\pm16$ & $46\pm12$ & $43\pm17$ \\
M~100    & 33 & $38\pm4$  & $57\pm13$ & $34\pm12$ & $46\pm12$ \\
\hline \hline
\end{tabular}}
\label{tab:fractions}
\end{center}
\end{table}

In Fig.~\ref{fig:heating_maps}, the left panel of each sub-figure shows the dust-heating map of the face-on view of each galaxy. The bulge and bar regions are denoted with a solid red circle and a dashed blue ellipse, respectively. The parameters used to define these regions were retrieved from the S$^4$G database \citep{2015ApJS..219....4S}. We define the bulge radius as $2 \times R_\text{e}$. \citet{2015ApJS..219....4S} used a modified Ferrers profile to model the bar. The histogram in the right panel of each sub-figure displays the young heating fraction distribution within the dust cells, weighted by the dust mass. For NGC~1365 we find that on average, 68\% of the dust heating (or dust emission) originates from the radiation produced by the young stellar populations. M~83 shows also a high $f_\text{young}$ with a mean value of 64\%, while in the cases of M~95 and M~100 the young and old stellar populations contribute approximately to one half each, with mean young heating fractions around 47\% and 57\%, respectively. However, in the case of M~95 the mode of the distribution is shifted to a much lower value ($\sim 37\%$) compared to the mean and median values. From the dust-heating maps of NGC~1365, M~83, and M~100 we can see that the star formation is for the most part concentrated in the spiral arms. In the case of M~95 the bulk of the dust is heated by the old stellar population, with few sites of star formation remaining in the circumnuclear ring and in the outer ring of molecular gas that surrounds the stellar bar.  

\begin{figure*}[t]
\centering
\includegraphics[width=16cm]{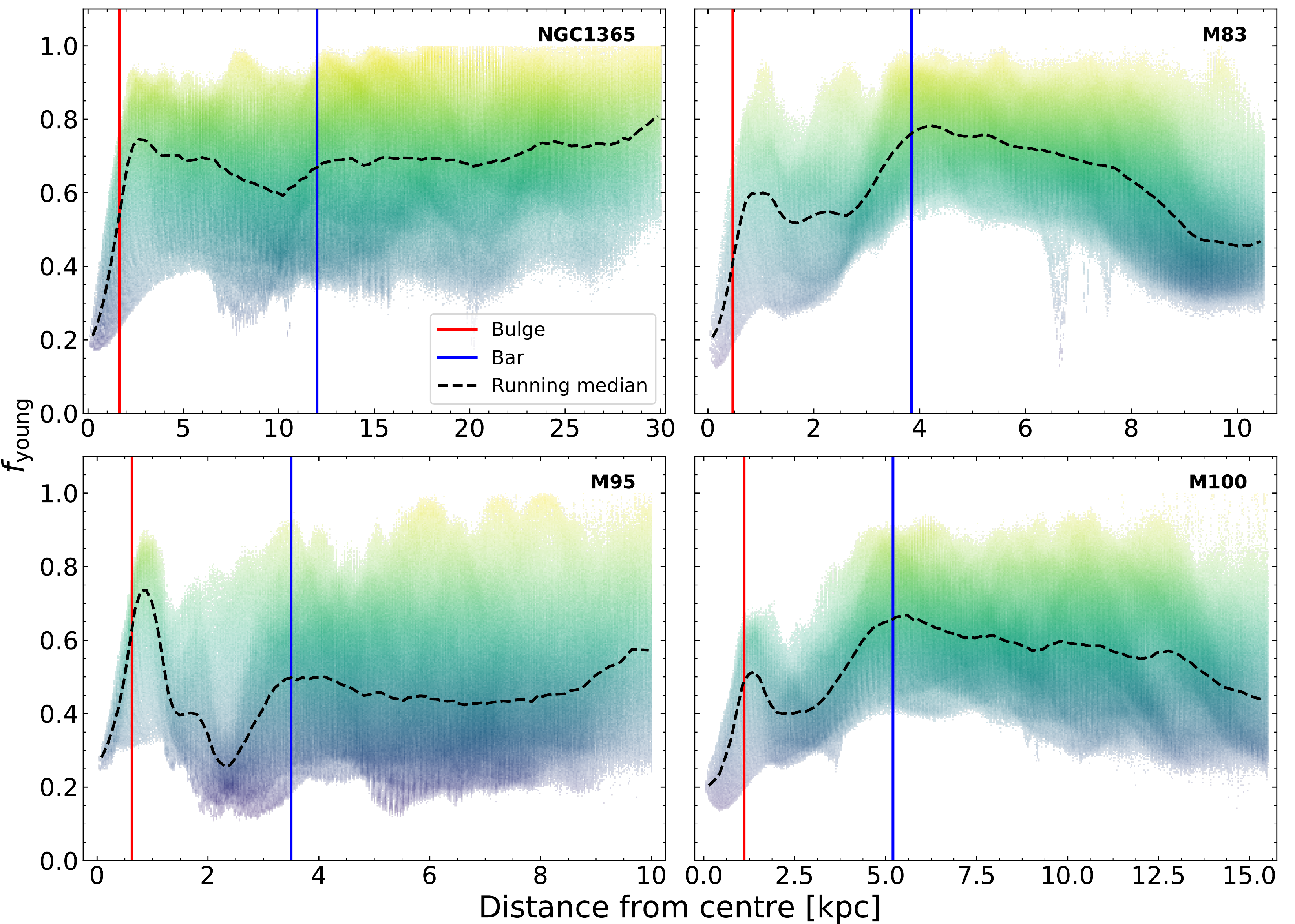}
\caption{Distribution of $f_\text{young}$ calculated for every galaxy with galactocentric distance. Each point represents a dust cell and it is colour-coded according to $f_\text{young}$. The level of transparency indicates the points density. The radius of the bulge is indicated with a vertical red line while the vertical blue line denotes the outer truncation radius of the Ferrers-bar. The dashed black line is the running median through the data points.}
\label{fig:rad_prof}
\end{figure*}

We find that the mean $f_\text{young}$ within the bulge region of every galaxy does not exceed $\sim 46\%$ (see Table~\ref{tab:fractions}). As expected, the old stellar population is the dominant dust-heating source in the central region of each galaxy \citep[see also,][]{2014A&A...571A..69D, 2017A&A...599A..64V, 2019_Verstocken}. Regarding the bar, although we do not treat the bar of each galaxy as a different component in our modelling, we can still extract basic information of its properties by looking at the young heating fraction, the radial profiles (see Fig.~\ref{fig:rad_prof}) and the dust temperature (see Sect.~\ref{subsec:Tdust}). In the bar region, the mean young heating fractions are: $\sim 66\%$, $\sim 59\%$, $\sim 43\%$, and $\sim 46\%$ for NGC~1365, M~83, M~95, and M~100 respectively, with the higher value being for the galaxy with the longest bar (NGC~1365; bar length of 24~kpc), while the lower value being for the galaxy with the shorter bar (M~95; bar length of 7~kpc). These fractions imply that the radiation field in the bar is caused by a mix of old and young stellar populations, both `equally' contributing to the dust heating. The young heating fractions for every galaxy and each region are given in Table~\ref{tab:fractions}.  

Furthermore, Fig.~\ref{fig:rad_prof} depicts the radial profiles of the young heating fractions. Each point represents a dust cell in our simulations. Following the running median (dashed black line) an interesting pattern appears. It is immediately evident, that all galaxies showcase a narrow central peak where the bulge region ends and the bar starts. This peak is followed by a local minimum in the young heating fraction and then an outer maximum which interestingly coincides with the bar truncation point. Then, the running median of $f_\text{young}$ in the galactic disc slowly declines or remains constant. This pattern is clearer for M~95 with a strong peak at 1~kpc distance, possibly due to the nuclear starburst and the inner star-forming ring that connects the bulge with the bar. Moreover, M~95 reaches a second minimum in the bar inner region indicating the suppression of star formation due to gas depletion or gas re-distribution. For example, \citet{2009A&A...501..207J} have shown the lack of H$\alpha$ emission in the bar region of M~95, and long-slit spectroscopy showed that any diffuse emission from that region is associated with post-AGB (Asymptotic Giant Branch) stars \citep{2015MNRAS.450.3503J}. In a recent study, \citet{2019A&A...621L...4G} presented evidence of suppressed star formation in the bar inner region of M~95 due to gas inflows to the nuclear region. On the other hand, the pattern we describe here is less prominent in the case of NGC~1365 possibly due to the enhanced star-forming activity close to the central area \citep{2019A&A...622A.128F}, but also due to the higher levels of star formation in the bar inner region. 

\begin{figure*}[t]
\centering
\includegraphics[width=15cm]{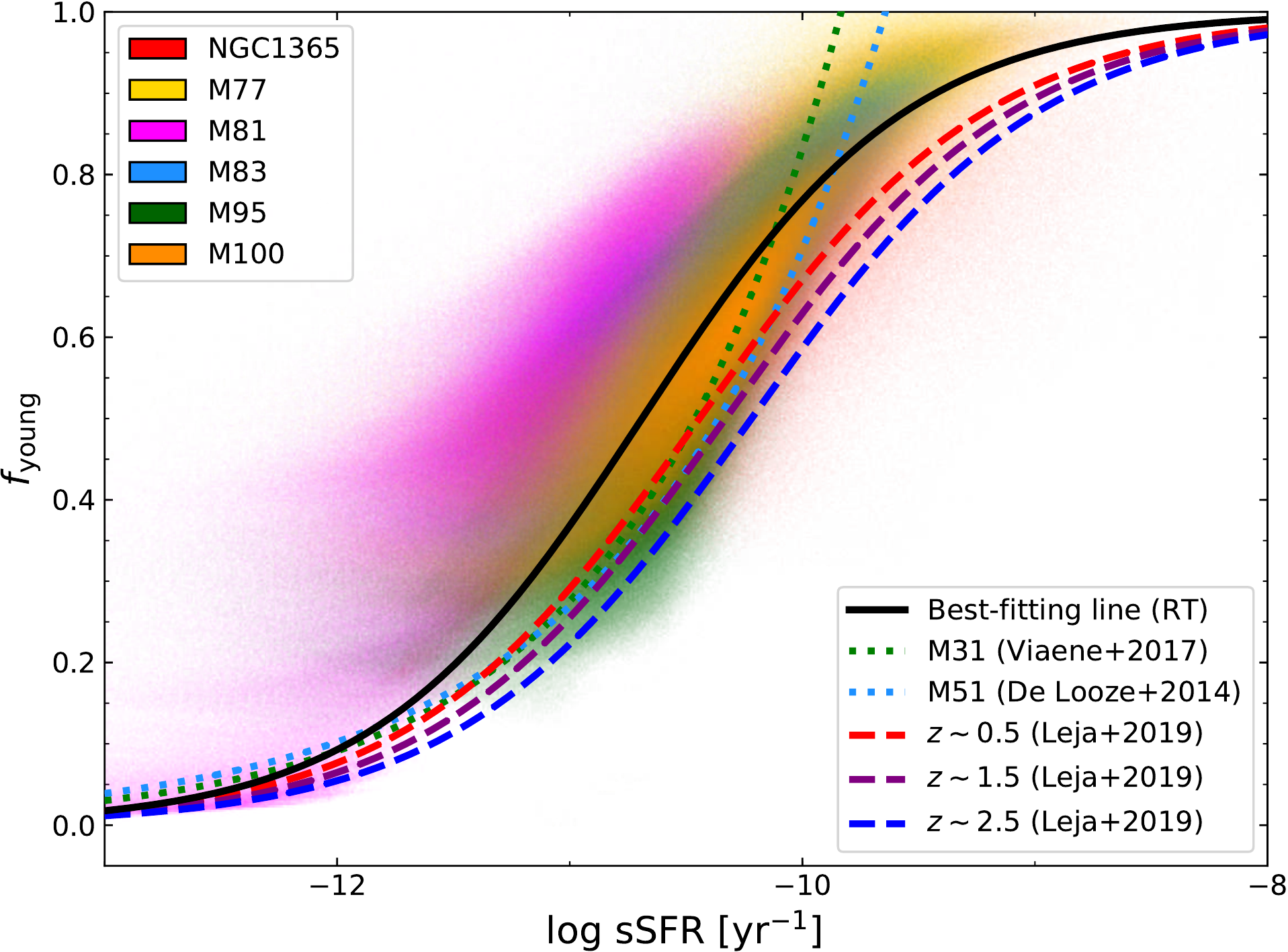}
\caption{Relation between sSFR and $f_\text{young}$, shown for the radiation transfer models of: NGC~1365, M~83, M~95, M~100 (this work); M~81 \citep{2019_Verstocken}; and M~77 \citep{Viaene2020}. Each galaxy dataset is assigned with a different colour indicated in the upper left corner of the figure. The solid black line shows the fit from Equation~\ref{eq:tanh} through all data cells of every galaxy (RT). For comparison purposes we also provide the best-fitted power-laws of: M~31 \citep[dotted green line,][]{2017A&A...599A..64V}; M~51 \citep[dotted cyan line,][]{2014A&A...571A..69D}; and the relations derived by \citet{2019ApJ...877..140L} for a sample of galaxies from the 3D-HST catalogues at three redshift bins: $z\sim0.5$ (dashed red line), $z\sim1.5$ (dashed purple line), and $z\sim2.5$ (dashed blue line).}
\label{fig:ssfr}
\end{figure*}

\citet{2009A&A...501..207J} reported the same pattern while studying the effects of bars on the radial distributions of H$\alpha$ and R-band light for more than 300 nearby galaxies. \citet{2009A&A...501..207J} have shown (see their Fig.~8) that the mean H$\alpha$ profiles (tracing the SFR) for galaxies with a clear optical bar and of Hubble stages $T$, between 3 and 5, have the same visible pattern as the one we observe here. They concluded as we do, that the bar component is responsible for the distinctive profiles seen in Fig.~\ref{fig:rad_prof}, since a similar pattern is absent in the radial profiles of unbarred galaxies (for example, see the radial profile of M~81; \citet{2019_Verstocken})

In summary, our analysis indicates that the central regions and the two diametrically opposed ends of the bar are places of enhanced star formation while the bar inner region is mostly populated by more evolved stars.  Even though the galaxy sample studied here is too small for any formal statistical analysis, we can confirm that bars have a clear effect on the variation of the $f_\text{young}$ radial profile. Of course, bars are not axisymmetric and therefore any effect caused by them will only arise in a diluted form, in any of the radial profiles. Furthermore, a dynamical origin of the presence of low-age stars in the bar central and outer regions cannot be excluded. \citet{2007A&A...465L...1W} have shown that the regions of enhanced star formation inside the bar are due to the accumulation of young stellar populations trapped on elliptical-like orbits along the bar. In any case, the distinct bar-induced features in the young heating fraction profiles suggest that the bars are prompting star formation that would not otherwise be happening \citep{2009A&A...501..207J}.

\subsection{Correlation between young heating fraction and sSFR} \label{subsec:ssfr}

In this section we report a strong relation between the young heating fraction and the sSFR. According to \citet{2014A&A...565A.128C}, sSFR is a measure of the hardness of the UV radiation field, providing an interesting and unique insight on the ratio of the current over the past star-forming activity of a galaxy. \citet{2014A&A...571A..69D} have shown the existence of a strong correlation between the sSFR and the dust mass fraction heated by the young stellar populations for M~51. This correlation was further confirmed in the radiation transfer models of M~31 \citep{2017A&A...599A..64V} and M~81 \citep{2019_Verstocken}. Radiation transfer models aside, others have found the same relationship at both local galaxy samples \citep{2016A&A...586A..13V, 2019A&A...624A..80N} and intermediate redshift ($z$) galaxies \citep{2019ApJ...877..140L}.

Figure~\ref{fig:ssfr} shows the relative contribution from the young stellar populations to the dust heating responsible for the TIR emission as was calculated for each dust cell and for all four galaxies of this study, as a function of $\log$~sSFR. Additionally, we include the data of the radiation transfer models of M~77 \citep{Viaene2020} and M~81 \citep{2019_Verstocken}. In total, the plot of Fig.~\ref{fig:ssfr} contains more than 15 million data points. To estimate the stellar masses we used the IRAC~3.6~$\mu$m luminosities and the prescription provided by \citet{2010MNRAS.405.2279O}. We converted the intrinsic FUV luminosity of the young stellar populations to SFR using the prescription provided in \citet{2012ARA&A..50..531K}. To calculate the sSFR in every data cell, we simply divided the SFR with the stellar mass. To fit the relationship of Fig.~\ref{fig:ssfr}, we used the function given in \citet{2019ApJ...877..140L} (see their Equation~2) which yields the following relation:

\begin{equation} \label{eq:tanh}
f_\text{young} = \frac{1}{2} \left[1 - \tanh\left(a \log\left(\text{sSFR}/\text{yr}^{-1}\right) + bz + c\right)\right] \, .
\end{equation}

\noindent where $a = -0.87$ and $c = -9.3$. Since the galaxies in our sample lie in the local Universe ($z < 0.01$) we used $z=0$. For comparison purposes, we provide the best-fitted power-laws of: M~31 \citep{2017A&A...599A..64V}, M~51 \citep{2014A&A...571A..69D}, and the relations derived by \citet{2019ApJ...877..140L} for a sample of galaxies from the 3D-HST catalogues at redshift $0.5 < z < 2.5$. Due to the overlap of colours, we present the data of the radiation transfer models of each galaxy separately in Fig.~\ref{fig:Fyoung_vs_sSFR_6}, and fit the data cells using both Equation~\ref{eq:tanh} and a power-law. The best-fitted parameters are given in Table~\ref{tab:Funev_sSFR}.

It is immediately evident that there is an increasing trend between the young heating fraction in each dust cell and the sSFR in all cases. Cells of high sSFR ($> 10^{-10}~\text{yr}^{-1}$) are primarily heated by the young stellar populations, whereas the contribution of the old population becomes more and more significant for cells with low sSFR ($\le 10^{-10}~\text{yr}^{-1}$). The bulk of data points of every galaxy are concentrated more or less in the same region of the diagram, with the sSFR spanning three orders of magnitude. Our results are in accordance with the relations produced by the radiation transfer models of M~31 and M~51 despite the overall differences and assumptions made in the studies of \citet{2017A&A...599A..64V} and \citet{2014A&A...571A..69D}, respectively (i.e. different ages of the young stellar populations and different methods of estimating the sSFR). The derived relationship, in principle, will enable us to quantify the young heating fraction based on sSFR measurements in other galaxies and can be applied to calibrate the energy fraction of the old stellar population in global SED modelling.

Furthermore, we find very good agreement with the relations derived by \citet{2019ApJ...877..140L} at different redshifts. The authors fitted the data of more than $\sim50,000$ galaxies from the 3D-HST catalogues at redshifts $0.5 < z < 2.5$. Galaxies at those redshifts are massive and obscured star formation is the main agent of star formation \citep{2017ApJ...838...19W}. The authors used the \textsc{Prospector-$\alpha$} physical model \citep{2017ApJ...837..170L} to fit the galaxy SEDs. The model includes a flexible non-parametric star-formation history (SFH), a two-component dust attenuation model with a flexible age-dependent \citet{2000ApJ...539..718C} attenuation curve, a model accounting for the MIR emission from AGN torii, and dust emission via energy balance. In their study, the young heating fraction is defined as the relevant fraction of $L_\text{UV+IR}$ emitted by the young stars ($\le 100~$Myr), while a \citet{2003PASP..115..763C} IMF was used. After fitting the data, the authors reported lower SFRs and higher stellar masses than those found by previous studies in the literature for galaxies at $0.5 < z < 2.5$. They infer that the cause for this offset in both quantities is the contribution from the old stars ($> 100~$Myr), implying an older, less active Universe. Here we notice that the relation yielded by Equation~\ref{eq:tanh} shifts towards higher sSFR values with increasing redshift (from the solid black line to the dashed blue line). It is also worth noting that for a fixed sSFR value the $f_\text{young}$ decreases with increasing redshift. To some degree the shift of the sSFR-$f_\text{young}$ relation towards higher sSFR values with increasing redshift can be attributed to the increased SFR, at least in the regime $0.5 < z < 1.5$. In addition, \citet{2019ApJ...877..140L} showed that the old stellar populations in high-redshifts ($1.5 < z < 2.5$) are relatively younger and on average more luminous, contributing more to the dust heating, which explains the decrease in $f_\text{young}$ with redshift.

The concluding remarks in \citet{2019ApJ...877..140L} agree quite well with the picture we draw here by studying the properties of local galaxies on resolved scales, as we also infer that the older stellar population has a more prominent role on the heating of the diffuse dust. The relation between the sSFR and the relative fraction of dust heated by the star-forming regions or by the old stellar populations has now been observed in a wide range of galaxy types and using various modelling approaches. Our analysis showcases the importance of a consistent modelling approach in order to derive safe conclusions when comparing different datasets. With that in mind, further investigation of the relationship discussed here, both in global and resolved scales, will allow for a better understanding of the scatter in the sSFR-$f_\text{young}$ relation.

\subsection{Dust temperature}\label{subsec:Tdust}

\begin{figure}[t]
\centering
\includegraphics[width=9cm]{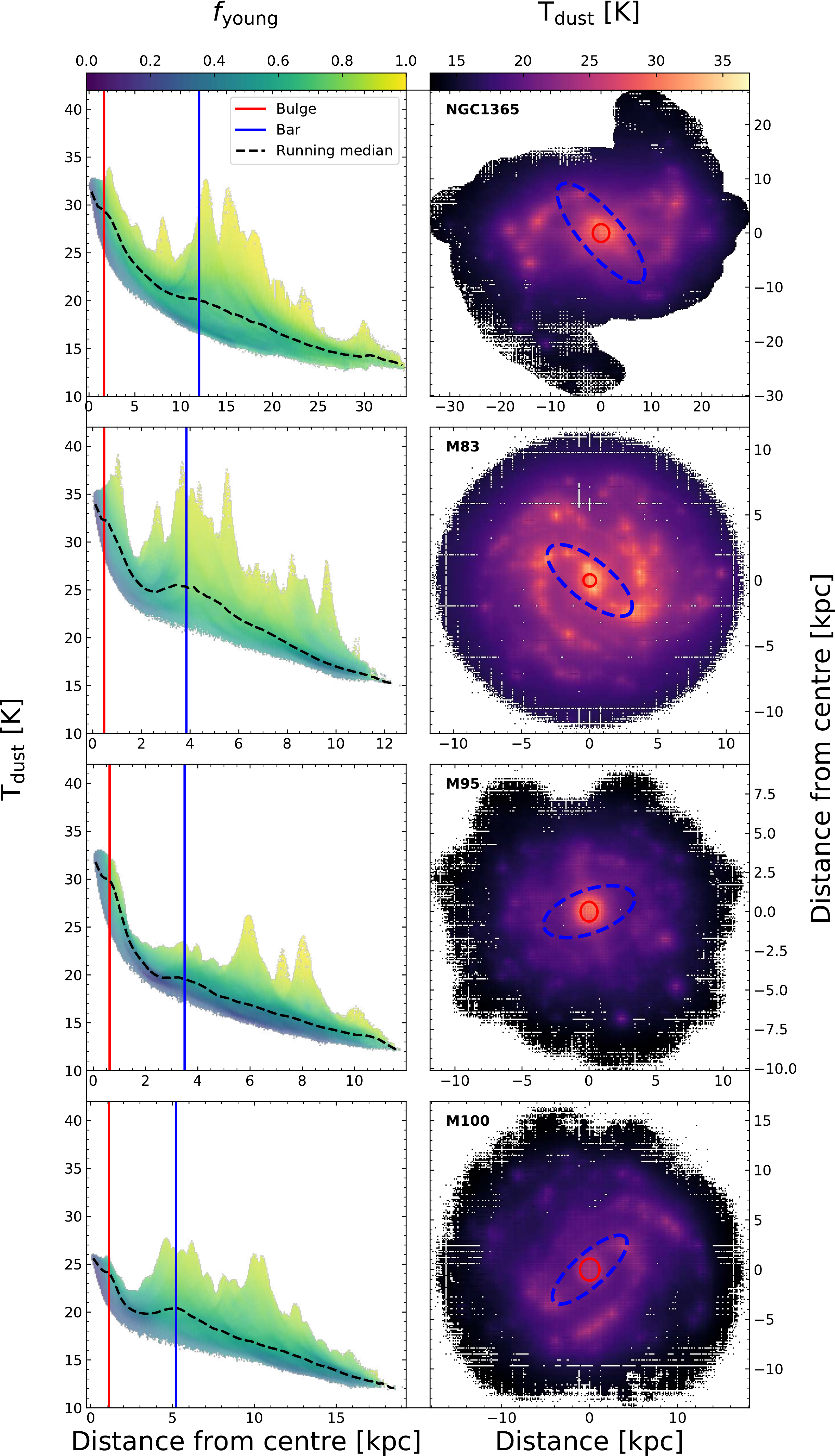}
\caption{Left column: distribution of the temperature of the diffuse dust with galactocentric distance. Each point represents a dust cell in our simulations. They are colour-coded according to $f_\text{young}$. The level of transparency indicates the point density. The bulge radius is indicated with a vertical red line while the vertical blue line denotes the outer truncation radius of the Ferrers-bar. The dashed black line is the running median through the data points. Right column: face-on view of the dust temperatures as obtained from the 3D dust cell distribution, for each galaxy. The bulge region is indicated with a solid red circle and the bar region with a dashed blue ellipse.}
\label{fig:Tdust}
\end{figure}

Light originating from star-forming regions acts as an important dust-heating source and thus one should expect to find a trend between regions of high dust temperatures and increased levels of star formation. Moreover, several studies have shown a dependence of the FIR surface brightness colours (i.e. indicators of dust temperature), with radius \citep{2010A&A...518L..65B, 2012MNRAS.419.1833B}. In the left column of Fig.~\ref{fig:Tdust} we plot the dust temperature ($T_\mathrm{dust}$) as a function of the deprojected galactocentric radius in kpc. Again, each point on each panel of this plot represents a dust cell in our simulations, colour-coded according to $f_\text{young}$. The bulge radius is indicated with a vertical red line while the outer truncation radius of the Ferrers-bar profile is indicated with the vertical blue line. From our analysis it is possible to determine how much the old stellar bulge and the composite stellar populations of the bar and disc structures affect the temperature of the diffuse dust.

Here we should make clear to the reader that the dust temperatures are only those of the diffuse dust and thus interpretation of the results should be considered with caution. Including the dense dust clouds in the star-forming regions, which they are subgrid properties of the MAPPINGS III templates, could add a significant amount of unusually high SFR and temperature values. 

The diffuse dust temperature of each dust cell in our simulations was approximated through the strength of the ISRF $(U)$. First we calculated $U$ by integrating the mean intensity of the radiation field $J_{\lambda}$ of each cell over the wavelength range 8-1000~$\mu$m. Then we normalised with the ISRF, estimated by \citet{1983A&A...128..212M} for the solar neighbourhood ($\sim 5\times 10^{-6}~$W/m$^2$). Assuming that dust is heated by an ISRF with a Milky-Way like spectrum \citep{1983A&A...128..212M} we employed the following equation to approximate the dust temperatures of the diffuse dust:

\begin{equation} \label{eq:Tdust}
T_\mathrm{dust} \ = T_\mathrm{o} \ U^{1/(4+\beta)} \, ,
\end{equation}   

\noindent \citep[][and references therein]{2019A&A...624A..80N}. Here, $T_\mathrm{o} = 18.3$~K which is the dust temperature measured in the solar neighbourhood, and $\beta$ is the dust emissivity index, which, for the \textsc{THEMIS} dust model, gives the value of 1.79.

Overall, temperatures range from 13-37~K with a decreasing trend towards the outermost regions of each galaxy, with several peaks and fluctuations which coincide with the appearance of high young heating fractions. These peaks are star-forming regions in the spiral arms or in the galactic disc, with the harsh UV radiation by the young populations heating the dust grains to high temperatures (25-37~K). Again, if we follow the running median line, a distinct pattern is seen in all galaxies. Dust temperature peaks at the centre of each galaxy and then sharply declines until a plateau is reached, approximately at 20-25~K (even with a rising trend, more clearly visible in the cases of M~83 and M~100). At the point where the bar is truncated, this plateau is followed by a continuous decline. These results are consistent with those obtained in previous studies on the $T_\mathrm{dust}$ radial trends of nearby spiral galaxies \citep{2010A&A...518L..72P, 2010A&A...518L..64S, 2011AJ....142..111B, 2012A&A...543A..74X, 2012MNRAS.419.1833B, 2012MNRAS.425..763G}. Here we see again the possible effect of the bar to the $T_\mathrm{dust}$ radial profile, since the plateau (or shoulder) is seen in the bar inner region. The average $T_\mathrm{dust}$ in the bulge region of NGC~1365, M~83, M~95, and M~100 is warmer than the average $T_\mathrm{dust}$ in the bar by 25\%, 22\%, 28\%, and 16\%, respectively. We measured the global dust temperatures for NGC~1365, M~83, M~95, and M~100 to be $19.2\pm3.8~$K, $21.8\pm3.6~$K, $17.5\pm3.0~$K, and $17.6\pm2.6~$K, respectively. The average dust temperature of our galaxy sample is $19.0\pm1.7~$K. According to \citet{2019A&A...624A..80N}, the average dust temperature for Sb-Sc type galaxies is $22.2\pm3.0~$K and compares fairly well with the mean dust temperature derived here for our galaxy sample.

Taking advantage of the information given by $f_\text{young}$, it is apparent that the emission of the old stellar population is directly responsible for the high dust temperatures at the nuclear region of each galaxy. This behaviour is expected since bulges are regions of extremely high radiation density produced by old stars. For example, several studies concluded that early-type galaxies (which tend to be more concentrated than spirals and their ISRF is governed by old stellar emission) have on average warmer dust temperatures than late-type galaxies \citep[e.g.][]{2011ApJ...738...89S, 2019A&A...624A..80N}. The old stellar population is also responsible for the lower dust temperatures in the bar and disc inner regions, as opposed to the higher temperatures there which are driven by star formation. In the right column of Fig.~\ref{fig:Tdust} we plot the dust temperature maps, to get a better visual view of the results discussed here. The bulge and the bar regions are indicated with a solid red circle and a dashed blue ellipse, respectively. Indeed, from this plot it is evident that the dust temperature is enhanced near the nucleus and along the spiral arms near star-forming regions. On the other hand, in the inter-arm and outermost regions of each galaxy, the diffuse dust is much colder. This radial trend mostly is a consequence of the diluted ISRF and possibly due to fewer young stellar populations at larger radii. The bars are not prominent in the temperature maps (with the exception of M~83). More specifically in the case of M~95, which has an inner and an outer star-forming ring with the bar acting as a bridge between them, we see that dust temperature in the inner ring ranges from 25-33~K, while the dust temperatures of the outer ring drops to 15-25~K. The old stellar population is the dominant heating agent of the diffuse dust in the outer ring of M~95, and the young stellar populations are dominating the dust-heating process in the inner ring.

\section{Conclusions} \label{sec:conclusions}

We have constructed detailed 3D radiative transfer models using the state-of-the-art Monte Carlo code \textsc{SKIRT}, for four late-type barred spiral galaxies (NGC~1365, M~83, M~95, M~100), with the purpose of investigating the dust-heating processes and to assess the influence of the bar on the heating fraction. Our models have been validated by comparing the simulated SEDs with the observational data across the entire UV to submm wavelength range, yielding a best-fitting description of each galaxy. Here we list our main results:

\begin{itemize}

\item We provide global attenuation curves for NGC~1365, M~83, M~95, M~100, M~81, and M~77, and we confirm the dependence of the shape of the observed attenuation curve with the star-to-dust geometry and the level of star-formation activity. The strength of the UV bump and the slope of the attenuation curve correlate with the sSFR of a galaxy and the degree of complexity of the star-to-dust geometry.

\item For the full sample, 36.5\% of the bolometric luminosity is absorbed by dust. This average fraction is in line with the mean values determined by \citet{2018A&A...620A.112B}, for the particular morphological group (Sb-Sc) that our galaxies fall into.

\item We find that the old stellar population has a more active role in the process of dust heating. This result hints that the use of infrared luminosity as proxy for the star-formation activity in star-forming galaxies should be used with caution. The global $f_\text{young}$ fractions for NGC~1365, M~83, M~95, and M~100 are 68\%, 64\%, 47\%, and 57\%, respectively. We find that the old stellar population is the dominant heating source in the bulge region, while both old and young stellar populations are equally responsible for the dust heating in the bar region.

\item We confirm a strong link between $f_\text{young}$ and the sSFR which was previously reported in the radiative transfer model analysis of M~51 \citep{2014A&A...571A..69D}, M~31 \citep{2017A&A...599A..64V}, and M~81 \citep{2019_Verstocken}, as well as in studies of \citep{2019A&A...624A..80N} for the DustPedia galaxy sample and \citet{2019ApJ...877..140L} for the 3D-HST galaxy sample, and provide a relation to calibrate the contribution of the old stellar population to dust heating in global SED modelling.

\item We confirm that the central regions and the two diametrically opposed ends of the bar are places of enhanced star formation and show that the bar in those galaxies affects the radial profiles of the $f_\text{young}$ and dust temperature. On average, the diffuse dust temperatures at the central regions of galaxies are warmer than those at the bar regions, while $T_\mathrm{dust}$ decreases towards the outer parts of galaxies. The old stellar population is exclusively responsible for the warmer $T_\mathrm{dust}$ at the bulge and the colder $T_\mathrm{dust}$ across the galactic disc of galaxies. The young stellar populations are responsible for the warmer $T_\mathrm{dust}$ in the spiral arms and near the star-forming dust clouds. The average dust temperature of our galaxy sample is $19.0\pm1.7~$K and is comparable to the mean values derived by \citet{2019A&A...624A..80N}, for the particular morphological group (Sb-Sc) that our galaxies fall into.

\end{itemize}

The full description of our framework and the results of the radiation transfer modelling of M~81 are presented in \citet{2019_Verstocken}, while the modelling results of a galaxy with the addition of an AGN component, NGC~1068 (M~77) will be presented in \citet{Viaene2020}. The continuation of the 3D radiation transfer modelling in a statistically significant sample of nearby spatially resolved galaxies, which have been modelled in a homogeneous way, will allow us to better understand the scatter in the sSFR-$f_\text{young}$ relation but also to investigate the properties of dust (e.g. composition, size distribution, etc.) and possible variations in the dust-heating processes among different galaxy types in the local Universe. 

\begin{acknowledgements}

We thank the referee, Adolf Witt, for his helpful comments and suggestions. DustPedia is a collaborative focused research project supported by the European Union under the Seventh Framework Programme (2007-2013) call (proposal no. 606847). The participating institutions are: Cardiff University, UK; National Observatory of Athens, Greece; Ghent University, Belgium; Universit\'{e} Paris Sud, France; National Institute for Astrophysics, Italy and CEA, France. AN gratefully acknowledges the financial support from Greece and the European Union (European Social Fund- ESF) through the Operational Programme «Human Resources Development, Education and Lifelong Learning» in the context of the project “Strengthening Human Resources Research Potential via Doctorate Research“ (MIS-5000432), implemented by the State Scholarships Foundation (IKY). This research made use of Astropy,\footnote{\url{http://www.astropy.org}} a community-developed core Python package for Astronomy \citep{2013A&A...558A..33A, 2018AJ....156..123A}. This research has made use of the NASA/IPAC Infrared Science Archive (IRSA; \url{http://irsa.ipac.caltech.edu}), and the NASA/IPAC Extragalactic Database (NED; \url{https://ned.ipac.caltech.edu}), both of which are operated by the Jet Propulsion Laboratory, California Institute of Technology, under contract with the National Aeronautics and Space Administration. The computational resources (Stevin Supercomputer Infrastructure) and services used in this work were provided by the VSC (Flemish Supercomputer Center), funded by Ghent University, FWO and the Flemish Government – department EWI.

\end{acknowledgements}

\bibliographystyle{aa}
\bibliography{References}
%
\appendix
%

\section{Global photometry}\label{ap:photometry}

Table~\ref{tab:phot} summarises the final aperture photometry flux densities extracted from the image data, used for the radiative transfer modelling. The bands that were not used in our modelling are indicated in boldface.


\begin{table*}[!ht]
\caption{Integrated flux densities for our galaxy sample in this paper, listed by increasing central wavelength. The bands not used in our modelling are indicated in boldface.}
\begin{center}
\scalebox{0.91}{
\begin{tabular}{lcccccc}
\hline 
\hline 
& & & NGC~1365 & M~83 & M~95 & M~100 \\
Instrument & Band & $\lambda_\mathrm{eff}$ & Flux density & Flux density & Flux density & Flux density \\
& & [$\mu$m] & [Jy] & [Jy] & [Jy] & [Jy] \\
\hline
GALEX       & FUV              & 0.154 & $0.043\pm 0.002$ & $0.287\pm 0.013$ & $0.016\pm 0.001$ & $0.031\pm 0.002$\\
GALEX       & NUV              & 0.227 & $0.061\pm 0.002$ & $0.470\pm 0.013$ & $0.027\pm 0.001$ & $0.054\pm 0.002$\\
SDSS        & u                & 0.359 & --               & --               & $0.081\pm 0.001$ & $0.202\pm 0.003$\\
SDSS        & g                & 0.464 & --               & --               & $0.340\pm 0.003$ & $0.498\pm 0.004$\\
SDSS/Other  & r/R$_\mathrm{C}$ & 0.612 & $0.907\pm 0.007$ & $5.455\pm 0.044$ & $0.632\pm 0.005$ & $0.815\pm 0.006$\\ 
SDSS        & i                & 0.744 & --               & --               & $0.896\pm 0.006$ & $1.134\pm 0.008$\\
SDSS        & z                & 0.890 & --               & --               & $1.007\pm 0.008$ & $1.237\pm 0.010$\\
2MASS       & J                & 1.235 & $1.796\pm 0.050$ & $9.818\pm 0.275$ & $1.470\pm 0.041$ & $1.681\pm 0.047$\\
2MASS       & H               & 1.662 & $1.710\pm 0.050$ & $11.870\pm 0.332$ & $1.946\pm 0.054$ & $2.117\pm 0.059$\\
2MASS       & Ks               & 2.159 & $1.698\pm 0.050$ & $8.350\pm 0.234$ & $1.573\pm 0.044$ & $1.430\pm 0.040$\\
WISE        & W1               & 3.352 & $1.214\pm 0.035$ & $6.113\pm 0.180$ & $0.814\pm 0.024$ & $0.922\pm 0.027$\\
IRAC        & I1               & 3.508 & $1.170\pm 0.035$ & $6.295\pm 0.190$ & $0.805\pm 0.024$ & $0.967\pm 0.030$\\
IRAC        & I2               & 4.437 & $0.884\pm 0.027$ & $4.124\pm 0.124$ & $0.498\pm 0.015$ & $0.628\pm 0.020$\\
WISE        & W2               & 4.603 & $0.885\pm 0.030$ & $3.818\pm 0.130$ & $0.439\pm 0.015$ & $0.519\pm 0.018$\\
IRAC        & I3              & 5.628 & $2.190\pm 0.066$ & $12.400\pm 0.370$ & $0.820\pm 0.025$ & $1.314\pm 0.040$\\
IRAC        & I4              & 7.589 & $5.210\pm 0.156$ & $30.051\pm 0.901$ & $1.612\pm 0.048$ & $3.318\pm 0.010$\\
WISE        & W3              & 11.56 & $4.164\pm 0.192$ & $21.105\pm 0.971$ & $1.080\pm 0.050$ & $2.452\pm 0.113$\\
WISE        & W4             & 22.09 & $12.472\pm 0.698$ & $45.804\pm 2.565$ & $2.690\pm 0.151$ & $3.690\pm 0.207$\\
MIPS        & 24              & 23.21 & $8.853\pm 0.443$ & $39.885\pm 1.994$ & $2.387\pm 0.119$ & $3.318\pm 0.166$\\
MIPS        & 70         & 68.44 & --                & $306.368\pm 30.640$ & --                & $35.647\pm 3.565$\\
PACS        & 70        & 68.92 & $138.496\pm 9.695$ & $448.555\pm 31.398$ & $25.907\pm 1.813$ & $42.932\pm 3.005$\\
PACS        & 100          & 100.8 & $214.973\pm 15.048$ & --              & $49.566\pm 3.470$ & $87.256\pm 6.108$\\
MIPS        & 160       & 152.6 & --                & $756.137\pm 90.740$ & --               & $117.714\pm 14.126$\\
PACS        & 160     & 153.9 & $204.472\pm 14.313$ & $834.000\pm 58.380$ & $54.741\pm 3.832$ & $115.215\pm 8.065$\\
SPIRE       & PSW        & 247.1 & $99.620\pm 5.480$ & $371.240\pm 20.420$ & $29.693\pm 1.633$ & $63.481\pm 3.491$\\
SPIRE       & PMW         & 346.7 & $43.280\pm 2.380$ & $148.972\pm 8.194$ & $13.183\pm 0.725$ & $26.801\pm 1.474$\\
\textbf{HFI}& \textbf{857} & 349.9 & $37.410\pm 2.390$ & $134.040\pm 8.578$ & $9.535\pm 0.610$ & $16.454\pm 1.053$\\
SPIRE       & PLW            & 496.1 & $15.085\pm 0.830$ & $50.356\pm 2.770$ & $4.804\pm 0.264$ & $9.054\pm 0.498$\\
\textbf{HFI}& \textbf{545}   & 550.1 & $11.470\pm 0.700$ & $34.851\pm 2.126$ & $2.544\pm 0.155$ & $4.753\pm 0.300$\\
\textbf{HFI}& \textbf{353}     & 849.3 & $2.424\pm 0.020$ & $3.900\pm 0.030$ & $0.766\pm 0.006$ & $1.056\pm 0.008$\\
\hline \hline
\end{tabular}}
\\[10pt]
\label{tab:phot}
\end{center}
\end{table*}

\newpage
\section{Image comparison}\label{ap:residual_maps}

\begin{figure*}[p]
\centering
\includegraphics[width=12.5cm]{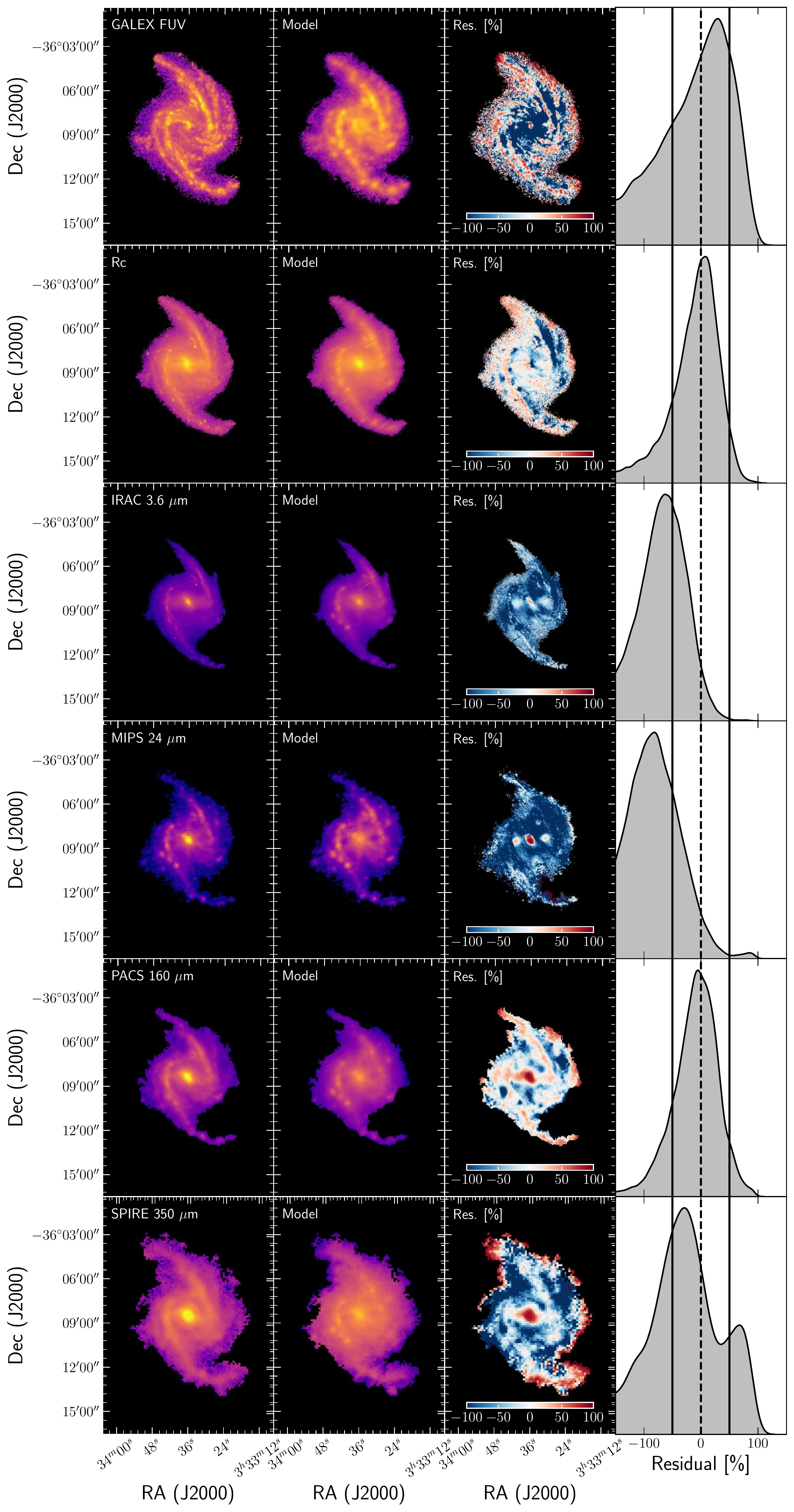}
\caption{Same as Fig.~\ref{fig:res_maps_m83} for NGC~1365}
\label{fig:res_maps_ngc1365}
\end{figure*}

\begin{figure*}[p]
\centering
\includegraphics[width=13.5cm]{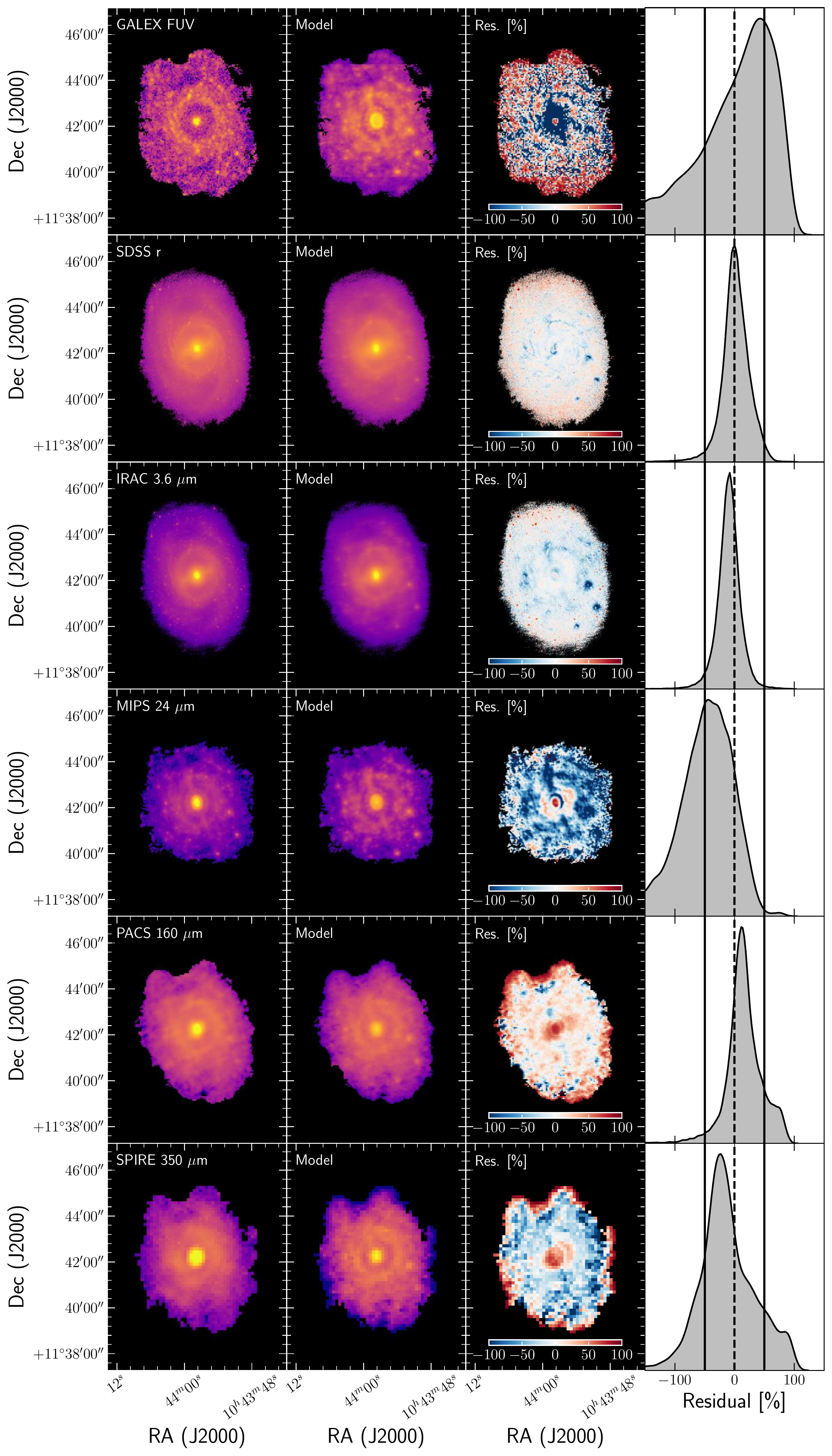}
\caption{Same as Fig.~\ref{fig:res_maps_m83} for M~95 but with the SDSS~$r$ observation used instead of $R_\mathrm{C}$.}
\label{fig:res_maps_m95}
\end{figure*}

\begin{figure*}[p]
\centering
\includegraphics[width=15cm]{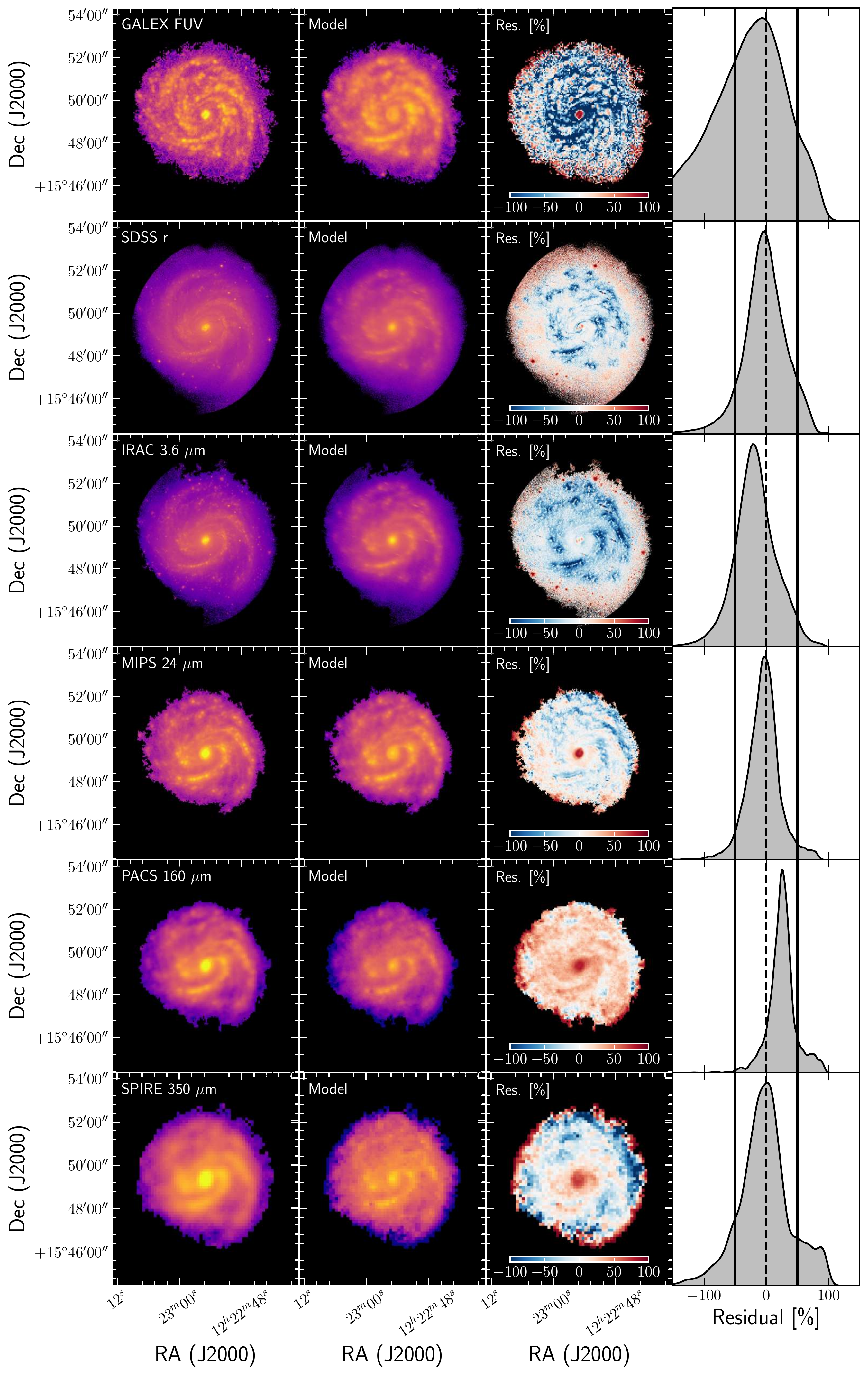}
\caption{Same as Fig.~\ref{fig:res_maps_m83} for M~100 but with the SDSS~$r$ observation used instead of $R_\mathrm{C}$.}
\label{fig:res_maps_m100}
\end{figure*}

Figures~\ref{fig:res_maps_ngc1365}, \ref{fig:res_maps_m95}, and \ref{fig:res_maps_m100} show the observational, model, and residual images for 6 wavebands that were fitted with \textsc{SKIRT}. Residuals are calculated as the relative difference between the modelled and the observed flux densities (Equation~\ref{eq:residuals}). Overall, the observations are fitted quite well with absolute residuals within 50\% in all three galaxies. The largest discrepancies can be seen for NGC~1365, for the IRAC~3.6$\mu$m and MIPS~24$\mu$m wavebands. The model overestimates the observations with absolute residuals higher than 50\%, especially for MIPS~24$\mu$m, where the model overestimates the flux densities up to 100\%, with the extremely bright AGN in the centre as a possible cause. In the fourth panel of Fig.~\ref{subfig:ngc1365_maps} (young ionising stellar disc), an Airy ring effect is still visible, despite our efforts to subtract the AGN emission (PSF) from the original image by employing 2D decomposition with \textsc{GALFIT} \citep{2010ApJ...721..193P}. To be more specific, since AGN is a point source we convolved it with the PSF for the MIPS~24$\mu$m image. We assumed a model for that galaxy that includes an AGN, a S\'{e}rsic bulge, a Ferrers bar and an exponential disc, and then we subtracted the modelled AGN from the original image. Nevertheless, the residuals of the remaining wavebands and galaxies are still more or less within 50\%, and with very narrow residual distributions, indicating that our simulations are accurate representations of the observed data. A detailed explanation of the cause of several residuals in these maps is given in Sect.~\ref{subsec:OvS}.

\newpage
\section{Correlation between young heating fraction and sSFR}\label{ap:Fyoung_vs_sSFR}

In this section we present the sSFR-$f_\text{young}$ relation for each galaxy and fit the bulk of the data cells using Equation~\ref{eq:tanh}, as well as a power-law. The best-fitting parameters of both fitting methods are given in Table~\ref{tab:Funev_sSFR} along with the Spearman's rank correlation coefficient ($\rho$). The strong correlation between the two quantities is justified by the fact that $\rho$ takes values $\ge0.80$. The only exception is NGC~1365 with $\rho=0.75$, however the correlation still remains strong. The best-fitting power-law for M~81 was given in \citet{2019_Verstocken}. An interesting result we notice here is that the slope of the power-law becomes more and more flat as the bulk of data cell values shifts towards higher sSFR and $f_\text{young}$ values. 

\begin{table}[h!]
\caption{The relationship in Fig.~\ref{fig:Fyoung_vs_sSFR_6} is fitted with a power-law: $y = a x + b$; where $y = \log f_\mathrm{young}$, $x = \log \left[\mathrm{sSFR}/\mathrm{yr}^{-1}\right]$, $a$ is the slope and $b$ is the intercept of the best-fitting line, and $\rho$ is the Spearman's rank correlation coefficient. The relationship is also fitted with Equation~\ref{eq:tanh}.}
\begin{center}
\scalebox{0.95}{
\begin{tabular}{l|ccc|cc}
\hline 
\hline 
\multirow{2}{*}{Galaxy ID} & \multicolumn{2}{c}{$y = a x + b$} & Spearman's coef. & \multicolumn{2}{c}{Equation~\ref{eq:tanh}} \\
\cline{2-6}
 & $a$ & $b$ & $\rho$ & $a$ & $c$ \\
\hline
NGC~1365 & 0.16 & 1.48 & 0.75 & -0.92 & -9.74 \\
M~77     & 0.12 & 1.13 & 0.84 & -0.80 & -8.75 \\
M~81     & 0.34 & 3.48 & 0.85 & -0.72 & -8.11 \\
M~83     & 0.22 & 2.06 & 0.85 & -0.96 & -10.2 \\
M~95     & 0.29 & 2.79 & 0.80 & -1.21 & -12.8 \\
M~100    & 0.22 & 2.04 & 0.84 & -0.83 & -8.86 \\
\hline \hline
\end{tabular}}
\label{tab:Funev_sSFR}
\end{center}
\end{table}

\begin{figure}[t]
\centering
\includegraphics[width=9cm]{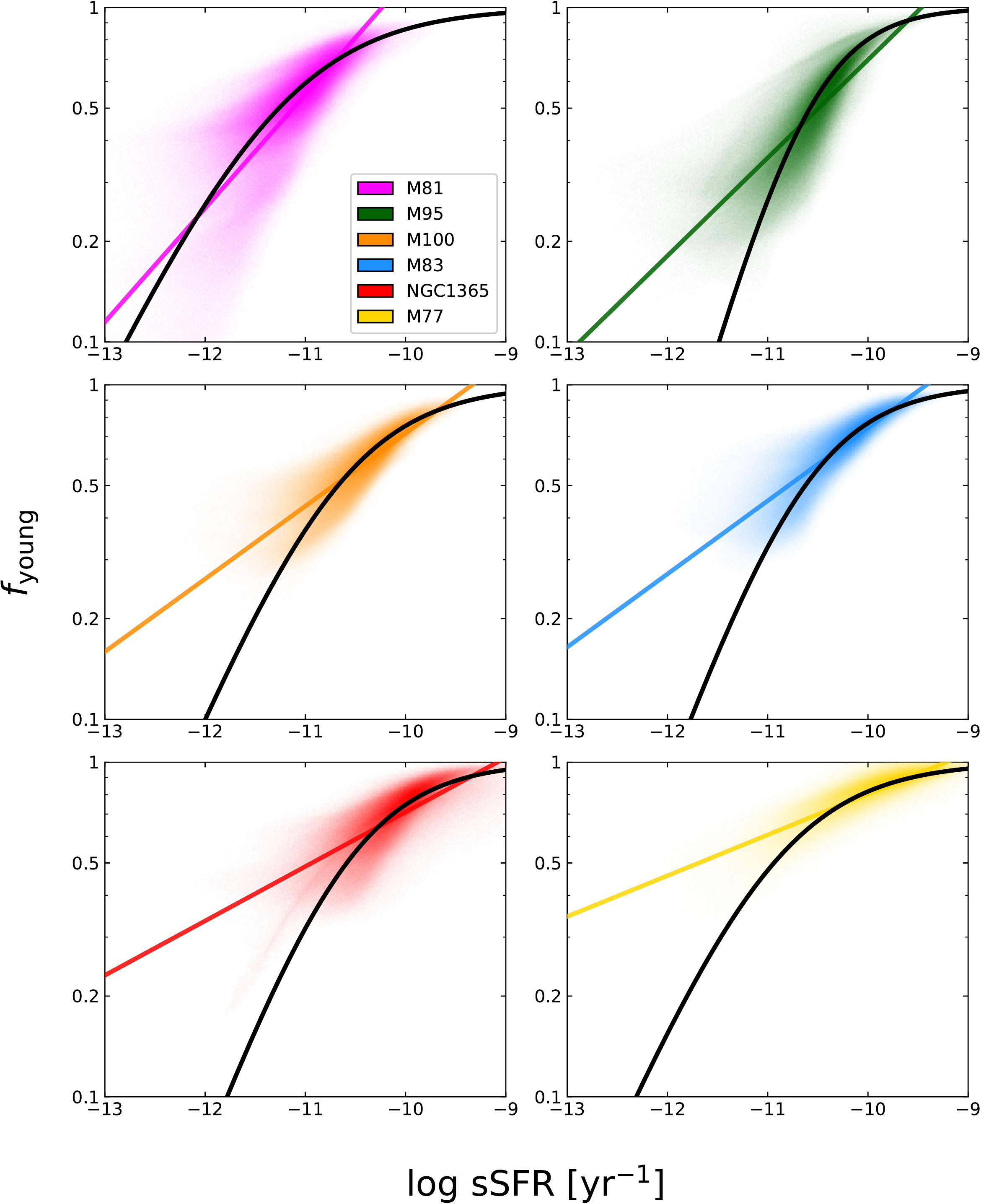}
\caption{Relation between sSFR and $f_\text{young}$, shown for the radiation transfer models of: NGC~1365, M~83, M~95, M~100 (this work); M~81 \citep{2019_Verstocken}; and M~77 \citep{Viaene2020}. Each galaxy dataset is assigned with a different colour indicated in the lower right corner of the first panel. The solid black line shows the fit from Equation~\ref{eq:tanh} through the bulk of data cells of every galaxy. Each coloured line shows the best-fitting power-law through the bulk of data cells of every galaxy.}
\label{fig:Fyoung_vs_sSFR_6}
\end{figure}


\end{document}